\documentclass[10pt,a4paper]{article}
\usepackage[utf8]{inputenc}
\usepackage{lmodern}
\usepackage[english]{babel}
\usepackage[a4paper, total={17.5cm, 23cm}]{geometry}
\usepackage{amsmath}
\usepackage{amssymb}
\usepackage{float}
\usepackage{graphicx}
\usepackage{hyperref}
\usepackage{setspace}

\usepackage[utf8]{inputenc}
\DeclareUnicodeCharacter{2009}{~}

\usepackage{siunitx}
\DeclareSIUnit\bar{bar}

\usepackage{xr}    

\usepackage[acronym, nopostdot]{glossaries}
\glsdisablehyper

\newacronym{2d}{2D}{two-dimensional}
\newacronym{hbn}{h-BN}{hexagonal boron nitride}
\newacronym{pmma}{PMMA}{poly(methyl methacrylate)}
\newacronym{cvd}{CVD}{chemical vapor deposition}
\newacronym{mbe}{MBE}{molecular beam epitaxy}
\newacronym{di}{DI}{deionized}
\newacronym{ipa}{IPA}{isopropyl alcohol}
\newacronym{maadf}{MAADF}{medium-angle annular dark field}
\newacronym{haadf}{HAADF}{high-angle annular dark field}
\newacronym{emccd}{EMCCD}{electron-multiplying charge coupled device}
\newacronym{eels}{EELS}{electron energy loss spectroscopy}
\newacronym{stem}{STEM}{scanning transmission electron microscopy}
\newacronym{uhv}{UHV}{ultra-high vacuum}
\newacronym{xps}{XPS}{X-ray photoelectron spectroscopy}
\newacronym{tof-sims}{TOF-SIMS}{time-of-flight secondary ion mass spectrometry}
\newacronym{afm}{AFM}{atomic force microscopy}
\newacronym{fov}{FOV}{field of view}
\newacronym{adf}{ADF}{annular dark field}
\newacronym{pid}{PID}{proportional-integral-derivative}

\preto\section{\setcounter{figure}{0}}

\makeatletter
\newcommand{\setFigureLabelForSection}{
  \ifnum\thesection=1
    \renewcommand{\figurename}{Figure}%
  \else
    \renewcommand{\figurename}{Supplementary Figure}%
    \renewcommand{\thefigure}{S\arabic{figure}}
  \fi
}
\makeatother

\preto\section{\setFigureLabelForSection}

\usepackage[style=numeric-comp, backend=biber, giveninits=true, minnames=3, maxnames=99, doi=true, isbn=false, sorting=none]{biblatex}  
\AtEveryBibitem{
  \clearfield{number}
  \clearfield{note}
  \clearfield{month}
  \clearlist{language}
}

\AtEveryBibitem{%
  \iffieldequalstr{entrykey}{irschik_raw_2025}
    {}
    {\clearfield{url}%
     \clearfield{urlyear}}%
}

\DeclareFieldFormat[article]{pages}{#1} 
\DeclareFieldFormat*{title}{#1} 
\DeclareFieldFormat*{journaltitle}{\mkbibemph{#1}} 
\DeclareFieldFormat[article,periodical]{volume}{\mkbibbold{#1}} 
\DeclareFieldFormat{urldate}{(accessed on  \thefield{urlday}\addspace  \mkbibmonth{\thefield{urlmonth}}\addspace \thefield{urlyear}\isdot)} 
\DeclareFieldFormat{doi}{DOI: \href{https://doi.org/#1}{\nolinkurl{#1}}} 
\DeclareFieldFormat{url}{URL: \href{}{\url{#1}}} 

\renewbibmacro*{publisher+location+date}{%
  \ifentrytype{book}
    {\printtext[parens]{\printlist{publisher}\setunit{\addcomma\space}\usebibmacro{date}}}}

\renewbibmacro*{in:}{%
  \ifentrytype{article}{}{\printtext{\bibstring{in}\intitlepunct}}}

\newcommand*{\mkbibbracketsuperscript}[1]{%
  \mkbibsuperscript{%
    \mkbibbrackets{#1}}}

\DeclareCiteCommand{\cite}[\mkbibbracketsuperscript]
  {\usebibmacro{cite:init}%

   \iffieldundef{prenote}
     {}
     {\BibliographyWarning{Ignoring prenote argument}}%
   \iffieldundef{postnote}
     {}
     {\BibliographyWarning{Ignoring postnote argument}}}
  {\usebibmacro{citeindex}%
   \usebibmacro{cite:comp}}
  {}
  {\usebibmacro{cite:dump}}

\renewcommand*{\familydefault}{\sfdefault}

\title{\textbf{Atomically clean free-standing two-dimensional materials through heating in ultra-high vacuum}}
\author{Philipp Irschik$^{1,2,*}$, David Lamprecht$^{1,3}$, Shrirang Chokappa$^{1,2}$, Clemens Mangler$^{1}$, Carsten Speckmann$^{1}$, Thuy An Bui$^{1,2}$, Manuel Längle$^{1}$, Lado Filipovic$^{3}$, Jani Kotakoski$^{1,*}$\\
$^1$University of Vienna, Faculty of Physics, Boltzmanngasse 5, 1090 Vienna, Austria\\
$^2$University of Vienna, Vienna Doctoral School in Physics, Boltzmanngasse 5, 1090 Vienna, Austria\\
$^3$Institute for Microelectronics, TU Wien, Gußhausstraße 27-29/E360, 1040 Vienna, Austria\\
$^*$Email: philipp.irschik@univie.ac.at, jani.kotakoski@univie.ac.at}
\date{\today}

\begin{document}
\baselineskip24pt

\begin{center}
    \LARGE\textbf{Atomically clean free-standing two-dimensional materials through heating in ultra-high vacuum} \\
\end{center}

\begin{center}
    \large
    Philipp Irschik$^{1,2,*}$, David Lamprecht$^{1,3}$, Shrirang Chokappa$^{1,2}$, Clemens Mangler$^{1}$, Carsten Speckmann$^{1}$, Thuy An Bui$^{1,2}$, Manuel Längle$^{1}$, Lado Filipovic$^{3}$, Jani Kotakoski$^{1,*}$\\
    $^1$University of Vienna, Faculty of Physics, Boltzmanngasse 5, 1090 Vienna, Austria\\
    $^2$University of Vienna, Vienna Doctoral School in Physics, Boltzmanngasse 5, 1090 Vienna, Austria\\
    $^3$Institute for Microelectronics, TU Wien, Gußhausstraße 27-29/E360, 1040 Vienna, Austria\\
    $^*$Email: philipp.irschik@univie.ac.at, jani.kotakoski@univie.ac.at \\
    \today
\end{center}


\begin{abstract}
\normalsize \bf \baselineskip22pt

Surface contamination not only influences but in some cases even dominates the measured properties of \acrlong{2d} materials. Although different cleaning methods are often used for contamination removal, commonly used spectroscopic cleanliness assessment methods can leave the level of achieved cleanliness ambiguous. Despite two decades of research on \acrshort{2d} materials, the true cleanliness of the used samples is often left open to interpretation. In this work, free-standing monolayer graphene and \acrlong{hbn} are annealed at different temperatures in a custom-built \acrlong{uhv} heating chamber, connected to a scanning transmission electron microscope via a vacuum transfer line, enabling atomically resolved cleanliness characterization as a function of annealing temperature, while eliminating the introduction of airborne contamination during sample transport. While annealing at 200~°C already reduces contamination significantly, it is not until 400~°C or higher, where over 90\% of the free-standing monolayer areas are atomically clean. At this point, further contamination removal is mainly limited by defects in the material and metal contamination introduced during the sample transfer or growth. The achieved large, atomically clean areas can then be used for further nanoscale engineering steps or device processing, facilitating interaction with the material rather than contamination.

\end{abstract} 

\baselineskip30pt

\newpage

\begin{refsection}

\subsection*{Introduction}

\Acrfull{2d} materials, such as monolayer graphene and \acrfull{hbn} possess a much larger surface-to-volume ratio compared to bulk materials, leading to an amplified effect of the often undesired surface contamination, which can result in the degradation of a variety of the materials' intrinsic properties, including their charge carrier mobility\cite{dan_intrinsic_2009, pirkle_effect_2011, suk_enhancement_2013}, wettability\cite{li_effect_2013}, and thermal conductivity\cite{pettes_influence_2011, jo_thermal_2013}. Additionally, the extent of its impact can vary between samples or even within different areas of the same specimen due to the lack of control over it, hindering reproducibility in device operations\cite{levesque_probing_2011}. Furthermore, even a thin film of adsorbed contaminants can obstruct the underlying material, with clean areas often being only a few to tens of nanometers in size\cite{lin_towards_2019}. This can be detrimental to contrast-based imaging techniques, such as (scanning) transmission electron microscopy, particularly for atomically thin free-standing specimens, where the signal of interest comes only from one to a few layers of atoms, and clean areas larger than a few nanometers are often desired\cite{trentino_atomic-level_2021}. These are necessary, as the observed properties may actually stem from interactions with contamination\cite{lee_measurement_2008, neumann_organic_2023, kohlrausch_one-size-fits-all_2025} rather than the material itself\cite{joudi_corrugation-dominated_2025, joudi_two-dimensional_2025}, and could lead to misinterpretation and misreporting of the properties of the studied materials.

Even though in-situ methods for sample preparation of 2D materials have been developed\cite{grubisic-cabo_situ_2023}, it is still common to use ex-situ procedures, such as mechanical exfoliation, molecular beam epitaxy, or \acrfull{cvd}, where the prepared samples are exposed to ambient conditions, either during or after the synthesis process\cite{whitener_graphene_2014, ji_chemical_2017, zhang_two_2017}, resulting in exposure to airborne hydrocarbon contamination, silicon, water vapor, and molecular oxygen already after a few minutes\cite{li_effect_2013, algara-siller_dry-cleaning_2014, susi_manipulating_2017, salim_airborne_2019, schweizer_mechanical_2020}. Also, transferring the materials onto other surfaces is typically needed for further device fabrication or applications, a process that often involves direct contact with polymers like \acrfull{pmma}\cite{reina_transferring_2008, jiao_creation_2008}. While organic solvents like acetone and chloroform\cite{cheng_toward_2011} can be used to dissolve \acrshort{pmma} after the transfer, they often fail to remove it completely, and can thus leave a thin film of polymer adsorbed on the surfaces\cite{lin_clean_2011, xie_clean_2015, cunge_dry_2015}, obstructing the underlying materials, and degrading their properties\cite{suk_enhancement_2013}.

This ubiquitous presence of surface contamination has incentivized the development of various post-preparation cleaning methods. While specialized cleaning techniques like cleaning in activated carbon\cite{algara-siller_dry-cleaning_2014, sun_force-engineered_2019}, electrolytic cleaning\cite{sun_characterization_2017}, or annealing with the aid of metal catalysts\cite{longchamp_ultraclean_2013, yulaev_toward_2016} have been explored, they are not suitable for all applications and often require specialized device fabrication. Thus, methods such as mechanical cleaning, plasma cleaning, laser cleaning, current-induced annealing, and thermal annealing are still often used\cite{zhuang_ways_2021}.

While mechanical cleaning methods can produce atomically clean areas by breaking the van der Waals bonds between the contamination and the underlying material using tips commonly used in \acrfull{afm}\cite{lindvall_cleaning_2012, schweizer_mechanical_2020}, this is an inherently slow process and thus does not scale well to larger areas. Plasma cleaning can be effective in etching surface-bound polymer contamination\cite{cunge_dry_2015}, however, a successful etching process relies on carefully chosen parameters for the plasma energy and density, as well as treatment duration, to not damage the underlying material in the process\cite{ferrah_xps_2016, ferrah_cf4h2_2019}. Cleaning techniques using laser ablation can also effectively remove contamination, but are limited to the laser spot size, restricting their application on the larger scale\cite{jia_toward_2016, tripathi_cleaning_2017}, and methods like current-induced annealing are limited to electrically conductive materials and require electrode attachment\cite{moser_current-induced_2007, bolotin_ultrahigh_2008, wang_ultraclean_2016}. Therefore, thermal solutions for sample cleaning are often used due to their rather simple installation and ease of use. Many setups allow annealing in controlled ambient environments, such as inert, reducing, and oxidizing atmospheres, as well as vacuum, the selection of which can have a drastic effect not only on the cleaning process but also on the integrity of the 2D material, with electrical doping\cite{ahn_procedure_2013, kumar_influence_2013}, defect formation\cite{lin_graphene_2012, gong_rapid_2013}, induced strain\cite{choi_effect_2015, wang_direct_2017}, and graphitization of carbonaceous contamination\cite{karlsson_graphene_2017} being common side effects of thermal annealing processes.

For graphene, even at temperatures as high as 500~°C, inert atmospheres have been reported to be ineffective at completely removing PMMA residue\cite{ahn_thermal_2016}, and can turn it into covalently bonded amorphous carbon, which makes it difficult to remove\cite{hong_origin_2013, sun_characterization_2017}. Reducing atmospheres can facilitate depolymerization of polymer residue when annealed at 400~°C or higher\cite{ahn_procedure_2013}, with H$_2$ and Ar/H$_2$ being particularly effective in removing residual conjugated carbon systems by transforming them into light hydrocarbons\cite{huang_characterization_2014}. However, this tends to induce rehybridization of graphene from \textit{sp}$^2$ to \textit{sp}$^3$, resulting in amorphous carbon\cite{lin_graphene_2012, choi_effect_2015}, as well as strong \textit{p}-type doping\cite{kumar_influence_2013}. Annealing in oxidizing atmospheres, while being effective in decomposing amorphous carbon contamination\cite{zhang_large-area_2019}, can cause the formation of cracks in graphene at temperatures as low as 200~°C\cite{lin_graphene_2012}, and still fails to remove all contaminants, even at 500~°C\cite{tripathi_cleaning_2017}. Also, it has been found that the oxidative strength of the atmosphere has a strong effect on the etching characteristics, as NO$_2$ seems to completely etch graphene away when annealed above 500~°C, while CO$_2$ preserves it\cite{gong_rapid_2013}. Thermal annealing in vacuum, while reported to be unable to remove contamination completely\cite{pirkle_effect_2011, siokou_surface_2011, xie_clean_2015, ahn_thermal_2016}, seems to introduce the least amount of lattice defects compared to other thermal annealing treatments due to the lack of reactive species in the annealing atmosphere\cite{cheng_toward_2011}. In addition, the amount of induced hole doping and strain is reported to be minimal\cite{kumar_influence_2013, wang_direct_2017}, while providing satisfactory cleaning results\cite{ahn_thermal_2016, tripathi_cleaning_2017}. The purity of the vacuum must be kept in mind though, as annealing under better vacuum conditions reportedly yields cleaner specimens\cite{hong_origin_2013}. Pressure values in the \SI{e-3}{\milli\bar} range have been shown to be ineffective at reliably cleaning graphene surfaces\cite{ahn_thermal_2016}, compared to annealing in high vacuum\cite{cheng_toward_2011} or \acrfull{uhv}\cite{pirkle_effect_2011, siokou_surface_2011, xie_clean_2015, wang_direct_2017}, where large atomically clean surfaces have been achieved\cite{tripathi_cleaning_2017}, making \acrshort{uhv} annealing a suitable candidate for contamination removal without harming the integrity of the underlying material.

Despite this established understanding of polymer degradation as well as hydrocarbon species desorption\cite{paserba_kinetics_2001, londero_desorption_2012} on graphene surfaces, there is still a lack of reporting on the cleaning of monolayer~\acrshort{hbn} despite its various projected applications, such as being a dielectric for nanoelectronic devices\cite{kim_synthesis_2012, okada_direct_2014, kim_synthesis_2015, jang_synthesis_2016, yan_direct_2015}, and for quantum information and sensor technology due to its quantum light emission across a wide spectral range\cite{bourrellier_bright_2016, tran_quantum_2016, gottscholl_initialization_2020, sajid_single-photon_2020, liu_single_2024}. Particularly, the latter relies on atomically clean surfaces to reliably conduct atomically-precise defect creation\cite{tran_robust_2016, grosso_tunable_2017, tang_structured-defect_2025} or carbon doping\cite{mackoit-sinkeviciene_carbon_2019, mendelson_identifying_2021, ren_atomistic_2025} to create active sites for single photon emission. While surface contamination reduction via thermal annealing of \acrshort{hbn} has previously been utilized, including annealing in oxidizing environments\cite{garcia_effective_2012, orofeo_ultrathin_2014, cartamil-bueno_mechanical_2017, byrne_atomic_2025}, and \acrshort{uhv}\cite{propst_automated_2024, paul_residue-free_2025}, the reports either lack information on the structure and abundance of contaminants as a function of annealing temperature, or do not employ atomically resolved characterization techniques, leaving open the possibility of a thin layer of contaminants remaining on the surface.

In this work, we investigate the effect of thermal annealing in UHV on CVD-grown and wet-transfer-assisted free-standing monolayer graphene and \acrshort{hbn}, using a custom-built UHV heating chamber that is connected to a scanning transmission electron microscope through a vacuum transfer line. This ensures that the cleaning effect can be directly characterized without introducing airborne contaminants in the process. Contrary to commonly used spectroscopic assessment methods with limited spatial resolutions, including Raman spectroscopy\cite{kumar_influence_2013, sun_characterization_2017}, \acrlong{xps}\cite{pirkle_effect_2011, ahn_procedure_2013, ahn_thermal_2016, ferrah_xps_2016}, and \acrlong{tof-sims}\cite{xie_clean_2015, wang_direct_2017}, we employ atomically resolved \acrfull{stem} to identify the presence of contamination on monolayer graphene and \acrshort{hbn} surfaces as a function of temperature down to the atomic level. Using \acrfull{adf}-STEM imaging, we can identify the position, shape, and abundance of contamination due to its higher scattering intensity compared to atomically clean free-standing monolayer regions. Using contrast-based image thresholding and segmentation to separate contaminated from clean areas, we determine the relative amount of surface area covered by contamination, a procedure that is carried out for different transfer methods and baking temperatures, correlating the relative clean monolayer area of graphene and \acrshort{hbn} with annealing temperature. This is complemented with \acrfull{eels} measurements before and after the annealing steps, giving insight into the composition of the contamination that survives the heating process, which is one of the main limitations for obtaining truly residue-free materials.

\subsection*{Methods}

\subsubsection*{Microscopy and spectroscopy}

All STEM measurements were carried out using a Nion UltraSTEM 100, an aberration-corrected scanning transmission electron microscope, operated at an acceleration voltage of 60 kV, with a beam convergence semiangle of 30 mrad and a probe size of approximately 1~\r{A}. The base pressure at the sample stage inside the microscope column was in the low \SI{e-9}{\milli\bar} range or lower at all times. The images were acquired using \acrfull{maadf} and \acrfull{haadf} detectors with collection semiangular ranges of 60-200 mrad and 80-300 mrad, respectively.

\acrshort{eels} measurements were carried out using a Gatan PEELS 666 spectrometer with an Andor iXon 897 \acrfull{emccd}. All EEL spectra were acquired with a total of 512 pixels and an energy dispersion of 0.5 eV/pixel at a collection semiangle of 35 mrad. All original spectra, including their background subtraction, are found in the Supporting Information.

\subsubsection*{Digital image processing}

MAADF-STEM images of graphene and \acrshort{hbn} that were acquired at a nominal \acrfull{fov} of 128×128~nm$^2$ (1024x1024~px, pixel time \SI{8}{\micro\second}) were subsequently subjected to a thresholding-based automated digital image processing algorithm to identify the relative abundance of clean monolayer regions in the frame. The lower thickness and thus lower scattering intensity of clean monolayer regions compared to contamination and multilayer structures in \acrshort{stem} enable contrast-based image thresholding and segmentation, with an example shown in Supplementary Figure~\ref{fig:thresholding_demonstration} (Supporting Information). For graphene samples heated at \SI{450}{\degreeCelsius}, images of entire holes (1024×1024~nm$^2$ FOV nominally, 4096×4096~px, pixel time \SI{8}{\micro\second}) in the sample support were acquired and thresholded instead, where a circular mask with a diameter of 950~nm was applied to mask the sample support area. We refer to the Supporting Information for a detailed description of the digital image processing steps.

\subsubsection*{Sample preparation}



{\bf Graphene}--- We used commercially available "Easy Transfer" monolayer graphene (Graphenea, Inc.), which is sandwiched between a \SI{100}{\micro\meter} thick porous water-releasable polymer support film and a sacrificial \acrshort{pmma} layer. It was cut into small (ca. 3×3~mm$^2$) squares roughly matching the size of a TEM grid. The graphene/PMMA stack was separated from the polymer support by dipping it into \acrfull{di}-water, making it float on the water's surface due to the hydrophobic nature of \acrshort{pmma}. The stack was then fished out with a TEM grid with a Quantifoil amorphous carbon support film (Quantifoil R 1.2/1.3 Au grid), after which it was left to dry in air on a standard laboratory hot plate at 150~°C for 1~h. It was then placed into a hot acetone (SigmaAldrich, ACS reagent, $\geq 99.5\%$) bath for 1~h to dissolve the majority of the \acrshort{pmma} residue, followed by a wash in \acrfull{ipa} (SigmaAldrich, ACS reagent, $\geq 99.5\%$) at room temperature for 45~min.

After investigating the sample quality and observing a large amount of metallic residue on all samples transferred this way, both before and after the annealing steps, further graphene samples were prepared via electrochemical delamination\cite{wang_electrochemical_2011} using commercial monolayer graphene on Cu with PMMA coating (Graphenea, Inc.). To delaminate the samples, after cutting them to size as described above, a platinum anode was placed into NaOH (SigmaAldrich, ACS reagent, $\geq 95.0\%$, pellets) aqueous solution (1~\textsc{m}), with the Cu/graphene/PMMA stack on Inox~02 tweezers acting as the cathode. All samples were delaminated at a voltage of 4~V. Afterwards, the delaminated graphene/PMMA stack was washed in \acrshort{di}-water to remove residual NaOH contamination, after which it was fished out with a TEM grid with a Quantifoil holey amorphous carbon membrane (Quantifoil R 1.2/1.3 Au grid), followed by a hot acetone bath for 1~h, with a final wash in \acrshort{ipa} for 45~min. One additional sample was transferred onto a holey silicon nitride support film (PELCO, \SI{3}{\milli\meter} diameter, \SI{200}{\micro\meter} silicon support structure, 0.5×0.5~mm window, 200~nm membrane thickness, \SI{1}{\micro\meter} pore size) using the same transferring steps.

{\bf \Acrlong{hbn}}--- After sourcing monolayer \acrshort{hbn} from three different suppliers, we chose monolayer \acrshort{hbn} on copper from SigmaAldrich due to their superior monolayer coverage (Figure~\ref{fig:hbn_quality_comparison}, Supporting Information), cut into small squares, which was spin-coated with \acrshort{pmma} (ALLRESIST AR-P 642.04, 200K) for 15 s at 500 RPM, followed by 15 s at 1000 RPM, and finally 60 s at 2000 RPM, at a ramp of 100 RPM/s. The Cu/\acrshort{hbn}/PMMA stack was placed on a glass slide, where the \acrshort{pmma} was left to cure on a hot plate at 150~°C for 30~min in air. After the \acrshort{pmma} had hardened, less than 1~mm was cut from all edges of the square to remove \acrshort{pmma} spillover to the other side of the foil. The subsequent delamination and washing steps were analogous to the graphene samples transferred via electrochemical delamination as described above.

\subsubsection*{UHV heating chamber}

All annealing experiments were carried out in a custom-built UHV heating chamber integrated into the CANVAS system at the University of Vienna\cite{mangler_materials_2022}, with a base pressure typically around \SI{5e-9}{\milli\bar} or lower. The samples are annealed via resistive heating of a tungsten filament that is placed out of the line of sight of the samples to avoid locally high temperatures and deposition of material coming off the wire. A more thorough description, including design sketches and used materials, can be found in the Supporting Information.

All samples were inserted at room temperature, after which the tungsten wire power was gradually increased to initiate the heating process, with typical heating rates between 0.1~°C/s and 1~°C/s, depending on the target temperature. The temperature was controlled by regulating the output current of the power supply using a \acrfull{pid} control algorithm. Upon reaching the desired temperature, the samples were kept at the setpoint for 3 h, after which the wire power was decreased to zero over five minutes. The samples were then kept in the chamber for about 30~min to passively cool down to temperatures between \qtyrange[range-phrase=~--~]{150}{250}{\degreeCelsius} (depending on the baking temperature) before being transferred to the microscope for characterization.

\subsection*{Results and Discussion}

We use commercial \acrshort{cvd}-grown monolayer graphene and \acrshort{hbn}, which were transferred onto TEM grids using a \acrshort{pmma}-assisted wet transfer process (see Methods). The majority of the \acrshort{pmma} residue was removed from the surface by placing the samples into an acetone bath, followed by a further washing step in \acrshort{ipa}. After preparation, the samples were inserted into the CANVAS system at the University of Vienna\cite{mangler_materials_2022}, which allows transporting the samples between a customized aberration-corrected scanning transmission electron microscope (Nion UltraSTEM 100)\cite{hotz_ultra-high_2016}, and a custom-built \acrshort{uhv} heating chamber without breaking the near-UHV conditions in between. This is integral to assessing the samples' cleanliness since this allows the cleaning effect to be directly characterized without intermediary air exposure. Upon insertion into the loadlock of the CANVAS system, they were subjected to a routine bake at 160~°C for over 10~h in high vacuum to remove water and reduce the amount of surface contamination to avoid contaminating the UHV system. Using the Nion UltraSTEM 100, all samples were investigated for coverage and cleanliness before the UHV annealing step, after which they were transferred to the UHV heating chamber and annealed at different temperatures for 3~h. After the baking procedure had finished, the samples were immediately returned to the microscope for post-heating characterization to minimize the effect of electron-beam-induced hydrocarbon deposition\cite{hugenschmidt_electron-beam-induced_2023}.

To assess the cleanliness of the samples, both before and after the thermal treatment, up to ten typical holes (\SI{1}{\micro\meter} nominal diameter) in the sample support were imaged.  The holes were chosen based on average and characteristic features resulting from each heating process. Different holes in the sample support were selected before and after each heating step, as the dissociation and subsequent cross-linking of the hydrocarbon contaminants from prior e-beam exposure interferes with heating-induced contamination removal\cite{dyck_mitigating_2017}. Approximately 20 arbitrarily chosen, non-overlapping areas with nominal FOV of 128×128~nm$^2$ were captured within each hole, acquiring a total of ca. 150 - 200 frames per temperature and sample. At each new frame position, the electron beam was manually re-focused onto the sample surface to obtain lattice resolution, minimizing out-of-focus features in the contamination at the interface to the material's surface. We note that no electron-beam-induced hydrocarbon deposition was observed during any measurement, and refer to the Supporting Information for a more thorough discussion. 

\begin{figure}[!t]
\centering
\includegraphics[width=\textwidth]{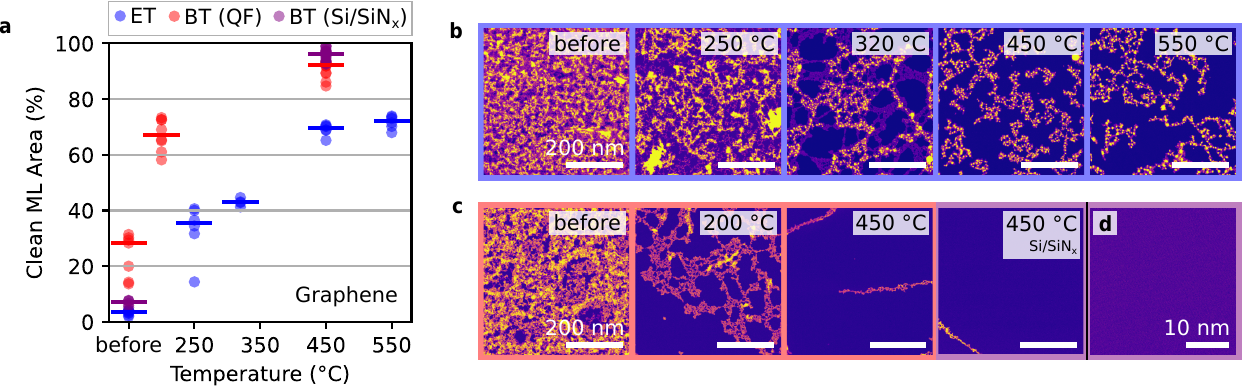}
\caption{{\bf Graphene cleanliness after annealing at different temperatures.} a) Relative clean monolayer (ML) area as a function of annealing temperature for Easy Transfer Graphene (ET, blue) on Quantifoil (QF), and graphene transferred via bubbling transfer (BT) on QF (red) and Si/SiN$_x$ (purple). Each dot (up to ten per temperature) represents the relative amount of clean monolayer area in an imaged hole in the sample support. The horizontal bars show the median. b) Cropped MAADF-STEM images of typical regions of free-standing monolayer Easy Transfer Graphene on QF before and after annealing at different temperatures. The darkest contrast is clean graphene, the brighter contrast is contamination. c) Graphene transferred onto QF and Si/SiN$_x$ via electrochemical delamination before and after annealing at different temperatures. d) Averaged MAADF-STEM image (50 frames) of large-area pristine monolayer graphene on Si/SiN$_x$ after heating at 450~°C for 2~h. Scale bars: b,c) 200~nm, d) 10~nm. The non-cropped images of the frames shown in panels b) and c) can be found in Supplementary Figure~\ref{fig:non-cropped_temperature_comparison}. Contrast has been adjusted to highlight relevant features, and we refer to Supplementary Figure~\ref{fig:colorbar_hole_images} for the false color grading reference.}
\label{fig:temperature_comparison_graphene_new}
\end{figure}

\subsubsection*{Easy Transfer Graphene}

All as-prepared Easy Transfer Graphene samples exhibited a high amount of metallic contamination in the form of nanosized iron and copper clusters in addition to carbonaceous contamination covering practically the entire surface. While atomically clean monolayer regions could be found, they were small in size, typically less than 10~nm in diameter, and accounted for less than 5\% of the free-standing regions (Figure~\ref{fig:temperature_comparison_graphene_new}a, blue dots). Example areas of contaminated and cleaned Easy Transfer Graphene are shown in Figure~\ref{fig:temperature_comparison_graphene_new}b. Heating to 250~°C did result in some removal of organic residue, increasing the average exposed clean area to 35\%. The sizes of the now uncovered clean areas remained small (around 20~-~30~nm in diameter), indicating that higher temperatures are needed. Heating to 320~°C uncovered more of the surface area (ca. 40\%), however, we identified thin hydrocarbon contamination networks all over the surface with only a marginal increase in the uncovered clean area compared to the 250~°C treatment. While annealing at 450~°C removed the majority of the carbonaceous contamination, resulting in a 70\% relative clean monolayer area, with clean regions wider than \SI{200}{\nano\meter} being common, we identified networks of metal clusters embedded in carbon-rich contamination as the primary residual contamination. Raising the annealing temperature to 550~°C did not increase the unveiled clean area, which remained at around 70\%. This not only indicates that the achievable cleanliness via \acrshort{uhv} annealing is heavily limited by the residual metal contaminants introduced during the sample growth and transfer process, but also that temperatures above 400~°C appear to be sufficient to remove almost all of the carbonaceous contamination that is not directly bound to metal clusters.

\subsubsection*{Graphene transferred via electrochemical delamination}

In comparison, Figure~\ref{fig:temperature_comparison_graphene_new}c shows that the graphene samples prepared via electrochemical delamination ("bubbling") exhibited a much higher amount of exposed clean monolayer area, found to be around 25\% (Figure~\ref{fig:temperature_comparison_graphene_new}a, red dots) as prepared, while still having the majority of the exposed surface covered by contamination. However, annealing at only 200~°C already resulted in a level of cleanliness rivaling the 450~°C treatment from the metal-contaminated samples (Easy Transfer), and revealed atomically clean areas thousands of~nm$^2$ in size. Raising the temperature to 450~°C removed almost all surface contamination, where, on average, over 90\% of the free-standing areas were identified to be atomically clean. Another sample that was transferred onto a holey Si/SiN$_x$ support and also annealed at 450~°C showed similar levels of cleanliness (Figure~\ref{fig:temperature_comparison_graphene_new}a, purple dots) as the ones transferred onto Au grids with a Quantifoil membrane. The main benefit of the rigid Si/SiN$_x$ support is the ability to obtain large area atomically resolved images (Figure~\ref{fig:temperature_comparison_graphene_new}d) without vibrations induced by the weakening of the sample support due to the bake, manifesting as non-ideal imaging conditions and impairing the acquisition of atomic resolution images. This is further discussed in the Supporting Information.

\begin{figure}[!t]
\centering
\includegraphics[width=\textwidth]{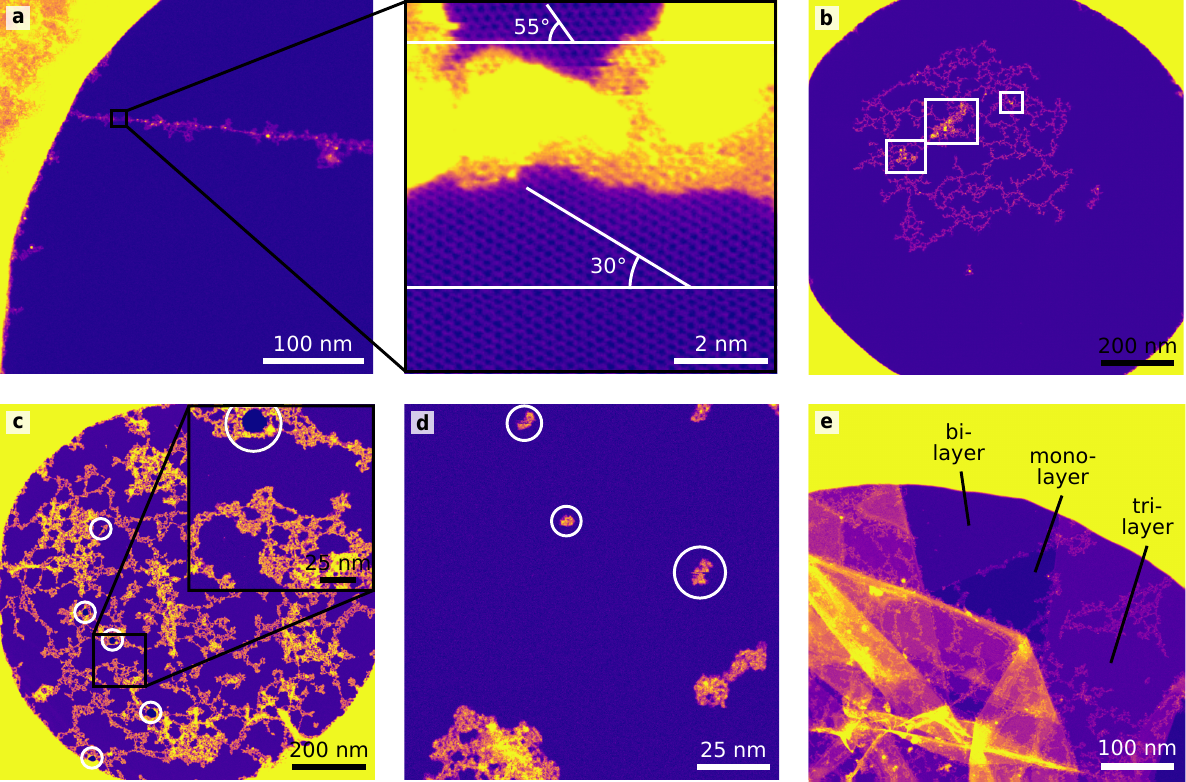}
\caption{{\bf MAADF-STEM images of the main cleaning limitations on graphene.} a) Grain boundary that is covered in hydrocarbon and metallic contamination, with large areas of clean graphene on both sides of the grain boundary. b) Aggregate of metal clusters (white rectangles) with dendritic hydrocarbon contamination anchored to it. c) Nanosized holes (white circles) with dendritic contamination structures around them. d) Small carbon-rich islands (white circles) embedded in what would otherwise be large clean areas. The observed streaking in the contamination islands is electron-beam-induced. e) Transfer-induced folds in graphene. The folded regions (lower left quadrant) are generally dirtier than mono- and bilayer regions, possibly due to a combination of their higher surface energy and potential contaminants trapped between the layers. The samples were annealed at a,b,e) 450~°C and c,d) 200~°C, respectively. Contrast has been adjusted to highlight relevant features, and we refer to Supplementary Figure~\ref{fig:colorbar_hole_images} for the false color grading reference.}
\label{fig:graphene_cleaning_limitations_new}
\end{figure}

\begin{figure}[hb]
\centering
\includegraphics[width=\textwidth]{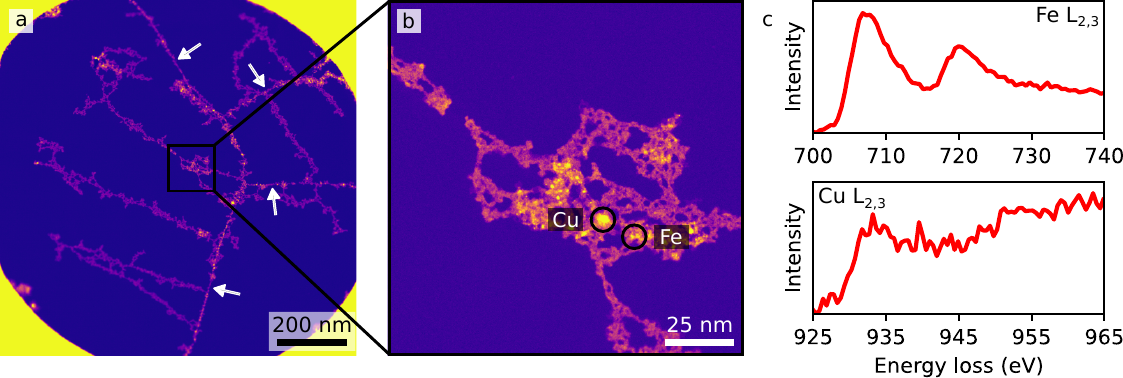}
\caption{{\bf Dendritic contamination structures emerging from ensembles of metal nanoclusters.} a) MAADF-STEM image of an example hole in the sample support with multiple grain boundaries (marked by white arrows) and a large amount of dendritic contamination structures. b) Higher magnification of the boxed area in a) showing an ensemble of Cu and Fe nanoclusters. c) Fe L$_{2,3}$ and Cu L$_{2,3}$ EEL spectra acquired inside the circles highlighted in panel b).}
\label{fig:dendritic_structures_new}
\end{figure}

The remaining residues after annealing at 450~°C consist mainly of copper, but also iron nanoclusters (Supplementary Figure~\ref{fig:metal_EELS_spectra_after_new}, Supporting Information), embedded in carbon-rich contamination, primarily located at graphene grain boundaries, shown in Figure~\ref{fig:graphene_cleaning_limitations_new}a. These metal clusters also form aggregates with a higher density of metallic residue, which are not necessarily attached to grain boundaries as seen in Figure~\ref{fig:graphene_cleaning_limitations_new}b, from which dendritic carbon-rich structures appear to emerge. Supplementary Figure~\ref{fig:dendritic_structures_new} (Supporting Information) shows that the metal contaminants are rich in Fe and Cu. We also identify nanosized holes in the material with typical sizes between 15 and 20~nm (Figure~\ref{fig:graphene_cleaning_limitations_new}c). However, based on the relatively low abundance of these holes, we presume that they were already present in the material before annealing and were covered by the contamination that was removed during the heating process, instead of being annealing-induced damage. Atomically clean areas that would otherwise be thousands of~nm$^2$ in size contain some isolated small carbon contamination patches (Figure~\ref{fig:graphene_cleaning_limitations_new}d), likely attached to point defects in the material. Contaminants in regions where graphene layers folded over on themselves during the sample transfer (Figure~\ref{fig:graphene_cleaning_limitations_new}e) are difficult to remove due to a rise in surface energy with increasing layer number\cite{kocherlakota_communication_2015}, which inhibits surface contamination desorption. Additionally, contaminants could not only be on the surfaces but also in between the layers, which would make them difficult to remove, as they cannot penetrate the layers they are sandwiched between. This results in typically dirtier areas compared to monolayer regions.

\subsubsection*{Hexagonal boron nitride}

\begin{figure}[!t]
\centering
\includegraphics[width=\textwidth]{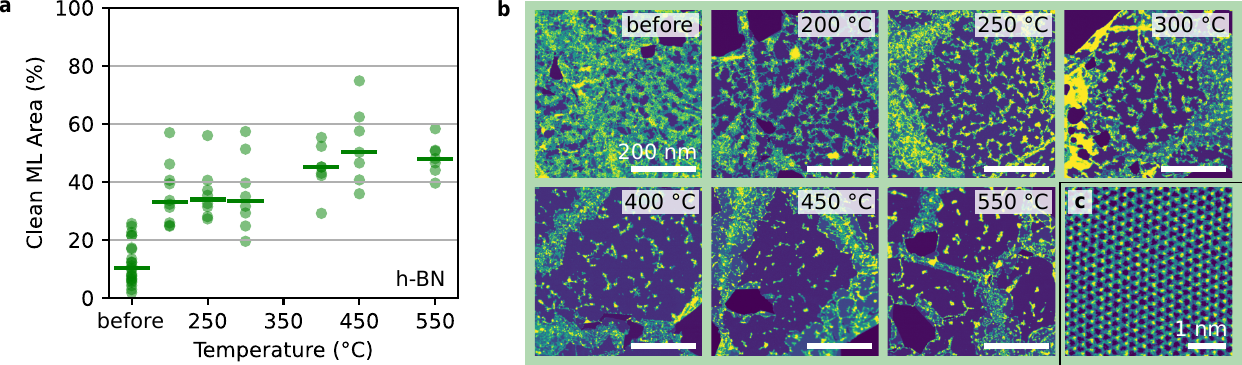}
\caption{{\bf \acrshort{hbn} cleanliness after annealing at different temperatures.} a) Relative clean monolayer (ML) area as a function of annealing temperature for h-BN. Each dot (up to ten per temperature) represents the relative amount of clean monolayer area in an imaged hole in the sample support. The horizontal bars show the median. b) Cropped MAADF-STEM images showing typical regions of free-standing monolayer \acrshort{hbn} before and after annealing at different temperatures. The darkest contrast is holes in the material, the second darkest contrast is clean monolayer \acrshort{hbn}, and the brighter features are multilayer structures and contamination. c) Gaussian filtered ($\sigma=3$) MAADF-STEM image of pristine monolayer h-BN after annealing at 200~°C. The non-ideal imaging conditions are attributed to residual heat and vibrations originating from heating-induced thinning of the sample support. Scale bars: b) 200~nm, c) 1~nm. The non-cropped images of the frames shown in panel b) can be found in Supplementary Figure~\ref{fig:non-cropped_temperature_comparison}. Contrast has been adjusted to highlight relevant features, and we refer to Supplementary Figure~\ref{fig:colorbar_hole_images} for the false color grading reference.}
\label{fig:temperature_comparison_hbn}
\end{figure}

\begin{figure}[hb]
\centering
\includegraphics[width=\textwidth]{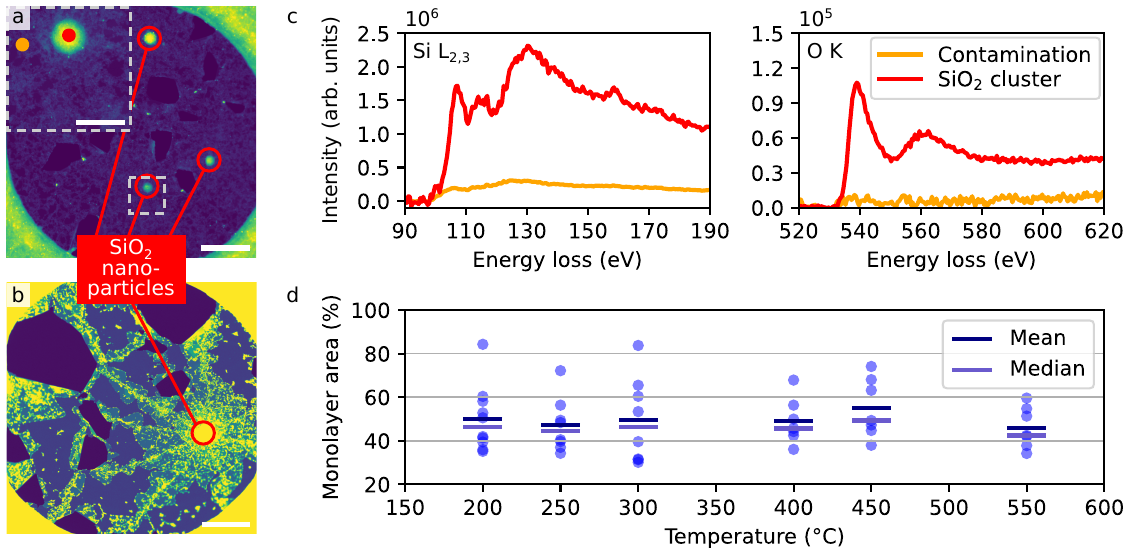}
\caption{{\bf SiO$_2$ nanoclusters on h-BN.} MAADF-STEM images of SiO$_2$ nanoparticles (marked with red circles) on a) dirty and b) cleaned \acrshort{hbn}. The inset in a) shows a selected particle at higher magnification. A different contrast enhancement in panel a) was chosen to highlight the morphologies of the spherical SiO$_2$ nanoparticles. The colored dots represent the positions where EEL spectra were acquired. c) Si L$_{2,3}$ and O K EEL spectra acquired on hydrocarbon contamination (orange) and a SiO$_2$ nanoparticle (red). d) Estimated h-BN monolayer coverage from different samples annealed at different temperatures. Each dot represents a different hole in the sample support. Scale bars: 200~nm (inset: 25~nm).}
\label{fig:SiO2_EELS_multilayer_new}
\end{figure}

\begin{figure}[hb]
\centering
\includegraphics[width=\textwidth]{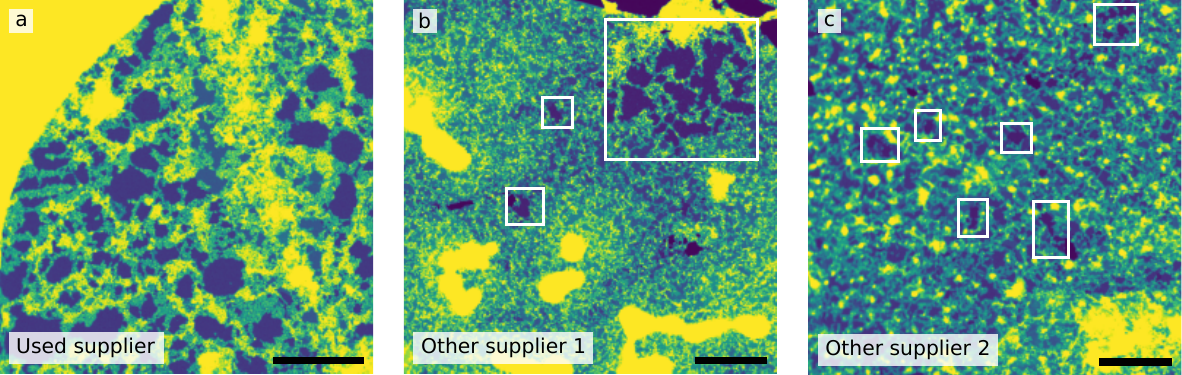}
\caption{{\bf MAADF-STEM images comparing the monolayer coverage from different suppliers.} a) Free-standing h-BN from the supplier used in this work. All areas with the darkest contrast are monolayer h-BN. Brighter features are bilayer regions, contamination, and the sample support (top left corner). The materials supplied by b) other supplier 1 and c) other supplier 2 exhibit a much lower amount of monolayer h-BN, where areas of notable sizes are highlighted with white rectangles. Scale bars: 100~nm.}
\label{fig:hbn_quality_comparison}
\end{figure}

Upon investigating the pre-cleaned samples, we observed that the quality of the transferred \acrshort{hbn} was inferior to that of graphene made via the same transfer method. Firstly, the suspended regions showed more holes and folds compared to graphene, likely originating from the transfer process. Secondly, a large fraction of the surface consisted of multilayer structures, which appeared to emerge from SiO$_2$ nanoclusters (Supplementary Figure~\ref{fig:SiO2_EELS_multilayer_new}, Supporting Information). We presume that they were introduced during the sample growth, possibly from evaporation of SiO$_2$ from the quartz tube during the \acrshort{cvd} growth process\cite{ruiz_silicon_2014}, rather than during storage or transfer. Together, these two sources contributed to an estimated monolayer coverage of only around 50\% (Supplementary Figure~\ref{fig:SiO2_EELS_multilayer_new}, Supporting Information), the highest among the three \acrshort{hbn} suppliers considered for this study (Supplementary Figure~\ref{fig:hbn_quality_comparison}, Supporting Information), compared to practically 100\% for graphene, despite the similar sample preparation procedures. This is also reflected in the strong variation of measured clean monolayer area (Figure~\ref{fig:temperature_comparison_hbn}a) across all temperatures, which is largely due to differences in monolayer coverage per imaged area, rather than local variations in residual contamination at a given temperature.

\begin{figure}[hb]
\centering
\includegraphics[width=\textwidth]{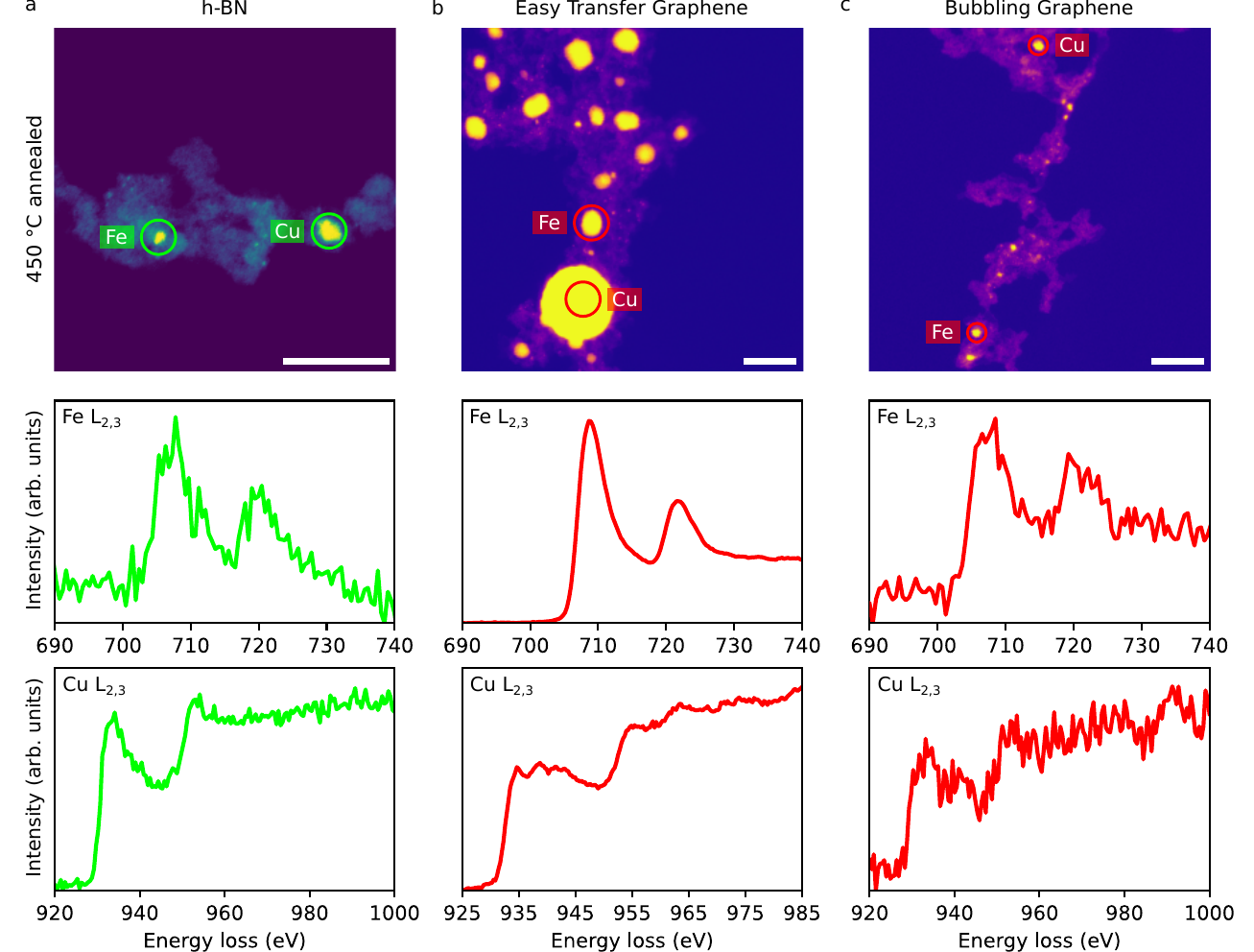}
\caption{{\bf EELS of metal contaminants.} Fe L$_{2,3}$ and Cu L$_{2,3}$ EEL spectra and accompanying MAADF-STEM (HAADF-STEM in panel a) images acquired on metallic contamination (colored circles) of a) h-BN, b) Easy Transfer Graphene, and c) graphene transferred via electrochemical delamination after annealing at 450~°C for 3~h. Scale bars: 10~nm.}
\label{fig:metal_EELS_spectra_after_new}
\end{figure}

\begin{figure}[hb]
\centering
\includegraphics[width=\textwidth]{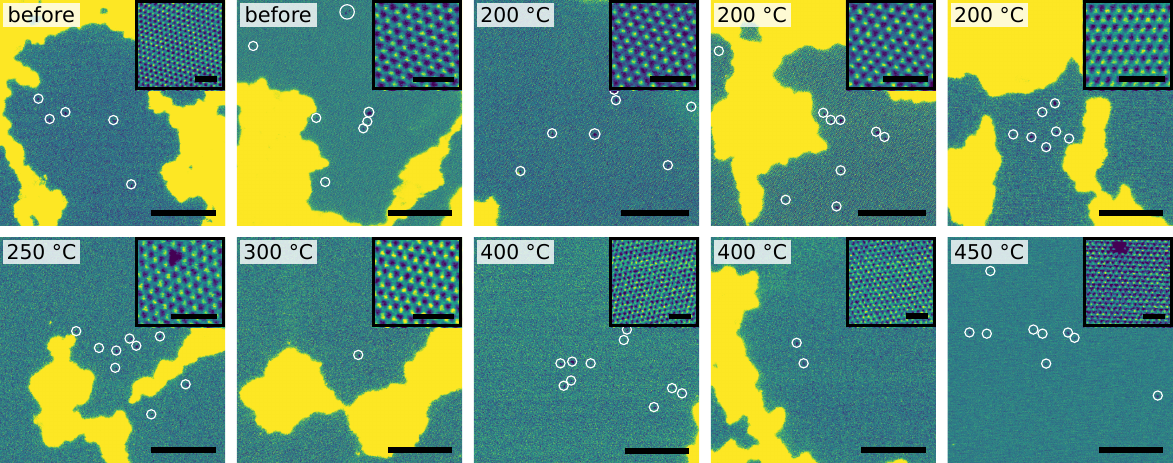}
\caption{{\bf Point defects in \acrshort{hbn}.} Examples of observed point defects in \acrshort{hbn} after annealing at different temperatures. Regions where vacancies could be unambiguously identified are marked with white circles. Images shown in the insets were taken towards the center of the corresponding frames. The higher defect density towards the center of each frame is due to prior e-beam exposure for the acquisition of the images in the insets. Scale bars: 10~nm (insets: 1~nm).}
\label{fig:hbn_defects}
\end{figure}

The as-prepared sample, where we measured an average clean area of only around 10\% (Figure~\ref{fig:temperature_comparison_hbn}a), was annealed at 200~°C (Figure~\ref{fig:temperature_comparison_hbn}b), after which around 30\% of the free-standing area was identified to be atomically clean monolayer \acrshort{hbn}. An example of atomically clean and defect-free \acrshort{hbn} annealed at 200~°C is shown in Figure~\ref{fig:temperature_comparison_hbn}c. In addition to contaminants adhered to the step edges between monolayer and bilayer regions, large portions of the surface are covered by networks of carbonaceous contamination with embedded clean areas in the range of only hundreds of~nm$^2$. These networks appeared to become more and more fragmented as the temperature was increased to 250~°C and 300~°C without substantially increasing the total clean area. Only when raising the temperature to 400~°C or higher did these networks disappear across all imaged suspended monolayer areas in the sample, and the fraction of the clean suspended monolayer area approached 50\%. We note that, accounting for the estimated relative free-standing monolayer area of ca. 50\%, the monolayer areas exhibit a cleanliness approaching 100\%, similar to what we found for graphene transferred via the same transfer method, annealed at the same temperatures. Similar to graphene, the residual observed contamination takes the form of nanosized islands, often but not always, around copper and iron nanoclusters (Supplementary Figure~\ref{fig:metal_EELS_spectra_after_new}). We also identify many small carbon-rich contamination agglomerations devoid of metallic residues within otherwise clean areas, which are likely attached to point defects in the material. However, the much higher density of these islands compared to what we found in graphene made via the same transfer method suggests a higher defect density of our transferred \acrshort{hbn}. While no systematic study was carried out here to characterize the defect density before or after the annealing steps, we did not observe a higher defect density in the atomically clean sites after annealing than before the treatment (Supplementary Figure~\ref{fig:hbn_defects}, Supporting Information), suggesting that UHV annealing can remove the majority of the surface contamination without harming the structural integrity of free-standing monolayer \acrshort{hbn} in the process. The higher defect density toward the center of the frames in Supplementary Figure~\ref{fig:hbn_defects} (Supporting Information) originates from electron-beam-induced damage\cite{bui_creation_2023} from prior imaging at higher local current densities to acquire the images in the insets. While it was straightforward to find atomically clean areas which were 200x200~nm$^2$ in size for graphene after the 200~°C bake, we observe a higher amount of residual contamination on \acrshort{hbn} that was annealed at the same temperature. The differences in the amount and morphology of contamination on \acrshort{hbn} and graphene after baking at only 200~°C further support a higher defect density of \acrshort{hbn}. While we did not investigate whether this is due to a lower quality of the source material or if the defects were introduced during sample transfer, recent literature also reported on a relatively high defect density (ca. \SI{0.03}{\per\nano\meter\tothe{2}}) in monolayer \acrshort{hbn} transferred via electrochemical delamination\cite{byrne_atomic_2025}. Given the high thermal stability of \acrshort{hbn}\cite{liu_ultrathin_2013, li_strong_2014}, and the lack of reactive species in our annealing atmosphere, in conjunction with only a modest annealing temperature, we presume that the 200~°C bake was not the source of our observed defects.

\subsubsection*{Recontaminated graphene sample}

\begin{figure}[!t]
\centering
\includegraphics[width=\textwidth]{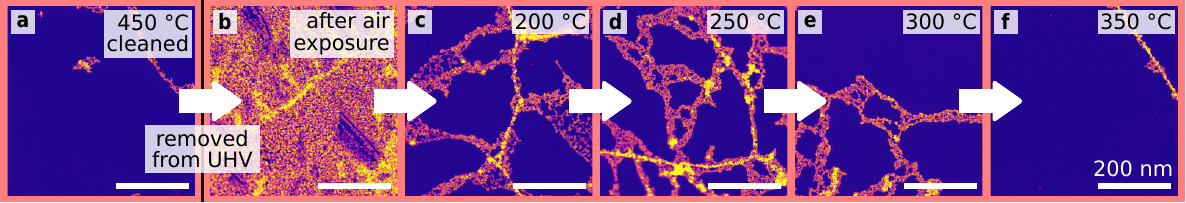}
\caption{{\bf Cropped MAADF-STEM images showing the heating response of a recontaminated sample.} a) Graphene sample prepared via electrochemical delamination that was cleaned at 450~°C for 3~h. b) The same sample, now full of contamination again, after removing it from UHV for 15~min followed by a 160~°C bake for over 10~h as part of the sample loading procedure. It was then iteratively heated at c) 200~°C, d) 250~°C, e) 300~°C, and f) 350~°C for 3~h each, until it exhibited the same level of cleanliness as before the recontamination step. In all images, the darkest contrast is clean monolayer graphene. Scale bars: 200~nm. The non-cropped images can be found in Supplementary Figure~\ref{fig:non-cropped_temperature_comparison}. Contrast has been adjusted to highlight relevant features, and we refer to Supplementary Figure~\ref{fig:colorbar_hole_images} for the false color grading reference.}
\label{fig:recontaminated_cleaning}
\end{figure}

To get a better understanding of the removal of only airborne (not polymer-induced) contamination, a cleaned graphene sample that was previously heated to 450~°C (Figure~\ref{fig:recontaminated_cleaning}a) was removed from the \acrshort{uhv} system for 15~min, during which it was exposed to ambient laboratory air, but not to polymers. After re-introducing it into \acrshort{uhv} through the regular ca. 160~°C vacuum bake overnight, the sample was found to again be heavily contaminated, with practically all of the surface being covered by hydrocarbon species (Figure~\ref{fig:recontaminated_cleaning}b). However, Figure~\ref{fig:recontaminated_cleaning}c shows that re-heating the sample to 200~°C already revealed more clean areas than the previous 200~°C bake (Figure~\ref{fig:temperature_comparison_graphene_new}c), where the contamination also consisted of \acrshort{pmma} residue. Then, the temperature was iteratively increased to higher values. While raising the temperature to 250~°C (Figure~\ref{fig:recontaminated_cleaning}d) did not increase the clean area, and 300~°C (Figure~\ref{fig:recontaminated_cleaning}e) provided only modest improvements, uncovering more and more of the pristine graphene lattice, it was only at 350~°C (Figure~\ref{fig:recontaminated_cleaning}f) that the sample exhibited the same level of cleanliness as after the previous 450~°C bake. This can be explained by the lack of required depolymerization of the \acrshort{pmma}, which is initiated by the breaking of the backbone of the \acrshort{pmma}, reportedly occurring between 350~°C and 400~°C\cite{kashiwagi_effects_1986, ahn_procedure_2013, xie_clean_2015, ahn_thermal_2016}, as the polymer residue had been removed by the previous 450~°C bake (before removal from \acrshort{uhv}), with mostly airborne hydrocarbons on the surface after the re-contamination step. 

\subsection*{Conclusions}

We report that thermal annealing in UHV is an effective method to remove airborne hydrocarbon and polymer contamination from both graphene and \acrshort{hbn} surfaces without compromising the integrity of the materials. Even annealing at temperatures as low as 200~°C can lead to noticeable contamination removal, especially when there is no polymer or metallic residue present, where well over 50\% of the probed free-standing monolayer surface is identified as atomically clean. This lower annealing temperature can be a viable cleaning step for fundamental study and manipulation of \acrshort{2d} materials, including impurity atom implantation and defect-engineering, where heating the specimen to higher temperatures might not be necessary or possible, and may only amplify the effect of the mismatch of the thermal expansion coefficient of the material and sample support, resulting in excess strain, induced corrugations, and tearing of the material. However, heating to higher temperatures is needed to achieve cleaner surfaces as the cleaning mechanism is incomplete until the materials are annealed at over 400~°C, where the decomposition and subsequent desorption of \acrshort{pmma} introduced during the sample transfer has occurred, and almost all of the contamination is removed. There, we find that achieving a higher degree of cleanliness is predominantly limited by defects in the material, such as grain boundaries and point defects, as well as residual metal clusters introduced during the sample growth and transfer processes. A high amount of metallic residue can be detrimental to the achievable cleanliness, as we found that a sample mostly devoid of metal clusters, when annealed at only 200~°C, can exhibit similar levels of cleanliness as a heavily metal-contaminated one that is baked at over 450~°C. Hence, we suggest to avoid using metal-based etchants during the sample transfer without further steps to dissolve the metallic residue, which \acrshort{uhv} annealing cannot remove. Also, when there are no polymer residues on the surface, we obtain cleaner surfaces at lower annealing temperatures, where heating graphene to 350~°C was sufficient in obtaining atomically clean areas thousands of~nm$^2$ in size, compared to over 400~°C if there was polymer present.

We also conclude that \acrshort{hbn} exhibits a thermal response to surface cleaning in UHV that is comparable to graphene. At temperatures above 400~°C, the surfaces are almost fully devoid of contamination, with contamination anchored to defect sites and metal clusters being the main limitation, similar to graphene. However, the quality of the material used in this work was found to be vastly inferior to that of graphene when samples transferred using the same transfer method were compared. Not only does the high abundance of multilayer structures originating from the growth heavily limit the amount of monolayer \acrshort{hbn}, but we also presume that the smaller average grain sizes of \acrshort{hbn} compared to graphene cause more damage to the material during the transfer. The resulting multilayer structures from the material folding over, covering up what would otherwise have been pristine monolayer regions, and inhibit contamination removal.

Finally, we observe only a minimal to no increase in defects for both graphene and \acrshort{hbn} after annealing, further reinforcing that UHV annealing can reliably clean surfaces without unintended defect creation. These large, clean surfaces can be beneficial for applications requiring large pristine areas, or for subsequent nanoscale engineering steps like defect-engineering and doping, all of which benefit from large exposed clean areas to not only increase the likelihood of interacting with the material rather than contamination, but to also utilize the improved physical properties of contamination-free \acrshort{2d} materials. Extending this cleaning procedure to other classes of \acrshort{2d} materials with different defect tolerances and less thermal stability, such as transition metal dichalcogenides\cite{pitthan_annealing_2019, kumar_direct_2020, tilmann_identification_2023}, is an important next step in evaluating its applicability for a broader range of materials, as we find that temperatures above 400~°C are necessary to remove polymer contamination from graphene and \acrshort{hbn}, resulting in nearly residue-free, pristine surfaces.

\subsection*{Acknowledgments}

We thank Lukas Eminger from Lithoz for his assistance with the design of the ceramic oven, and Lithoz for providing the 3D-printed oven components. This research was funded in part by the Austrian Science Fund (FWF) [10.55776/COE5, 10.55776/DOC142, 10.55776/P35318]. For open-access purposes, the author has applied a CC-BY public copyright license to any author-accepted manuscript version arising from this submission.

\subsection*{Author contribution}

P.I., D.L., and C.M. designed the UHV heating chamber, with further contribution from C.S. and M.L. when integrating it into the UHV system. S.C. and P.I. developed and integrated the sample transfer process, with P.I. also preparing the samples. T.A.B. prepared further hexagonal boron nitride samples used for testing the UHV heating stage. P.I. carried out the imaging for all samples. P.I. and D.L. conceptualized the data analysis methodology, with P.I. conducting the data analysis. J.K. and L.F. supervised the study. The manuscript text was initially written by P.I., and improved upon by all authors.

\subsection*{Data availability statement}

The experimental data supporting these ﬁndings will be openly available at PHAIDRA at https://phaidra.univie.ac.at/o:2149991 upon publication\cite{irschik_raw_2025}.

\subsection*{Supporting Information}

See Supporting Information for additional discussion regarding electron-beam-induced hydrocarbon deposition, sample support comparison, applied false color grading of ADF-STEM images, and a thorough description of the digital image processing steps. Additional figures on the MAADF-STEM image acquisition process,  EEL spectra of metal and SiO$_2$ contaminants, \acrshort{hbn} supplier comparison and defect observations, and design sketches of the \acrshort{uhv} heating chamber are also provided.

\newpage

\printbibliography

\end{refsection}


\newpage

\begin{refsection}

\section*{}

\renewcommand*{\familydefault}{\sfdefault}
\title{Supporting Information for}
\title{\textbf{Atomically clean free-standing two-dimensional materials through heating in ultra-high vacuum}}
\author{Philipp Irschik$^{1,2,*}$, David Lamprecht$^{1,3}$, Shrirang Chokappa$^{1,2}$, Clemens Mangler$^{1}$, Carsten Speckmann$^{1}$, Thuy An Bui$^{1,2}$, Manuel Längle$^{1}$, Lado Filipovic$^{3}$, Jani Kotakoski$^{1,*}$\\
$^1$University of Vienna, Faculty of Physics, Boltzmanngasse 5, 1090 Vienna, Austria\\
$^2$University of Vienna, Vienna Doctoral School in Physics, Boltzmanngasse 5, 1090 Vienna, Austria\\
$^3$Institute for Microelectronics, TU Wien, Gußhausstraße 27-29/E360, 1040 Vienna, Austria\\
$^*$Email: philipp.irschik@univie.ac.at, jani.kotakoski@univie.ac.at}
\date{\today}

\null
\vfill
\begin{center}
    \LARGE
    Supporting Information for \\
    \textbf{Atomically clean free-standing two-dimensional materials through heating in ultra-high vacuum} \\
\end{center}

\begin{center}
    \large
    Philipp Irschik$^{1,2,*}$, David Lamprecht$^{1,3}$, Shrirang Chokappa$^{1,2}$, Clemens Mangler$^{1}$, Carsten Speckmann$^{1}$, Thuy An Bui$^{1,2}$, Manuel Längle$^{1}$, Lado Filipovic$^{3}$, Jani Kotakoski$^{1,*}$\\
    $^1$University of Vienna, Faculty of Physics, Boltzmanngasse 5, 1090 Vienna, Austria\\
    $^2$University of Vienna, Vienna Doctoral School in Physics, Boltzmanngasse 5, 1090 Vienna, Austria\\
    $^3$Institute for Microelectronics, TU Wien, Gußhausstraße 27-29/E360, 1040 Vienna, Austria\\
    $^*$Email: philipp.irschik@univie.ac.at, jani.kotakoski@univie.ac.at \\
    \today
\end{center}
\vfill




\newpage

\subsection*{Applied color maps}

\begin{figure}[!t]
\centering
\includegraphics[width=\textwidth]{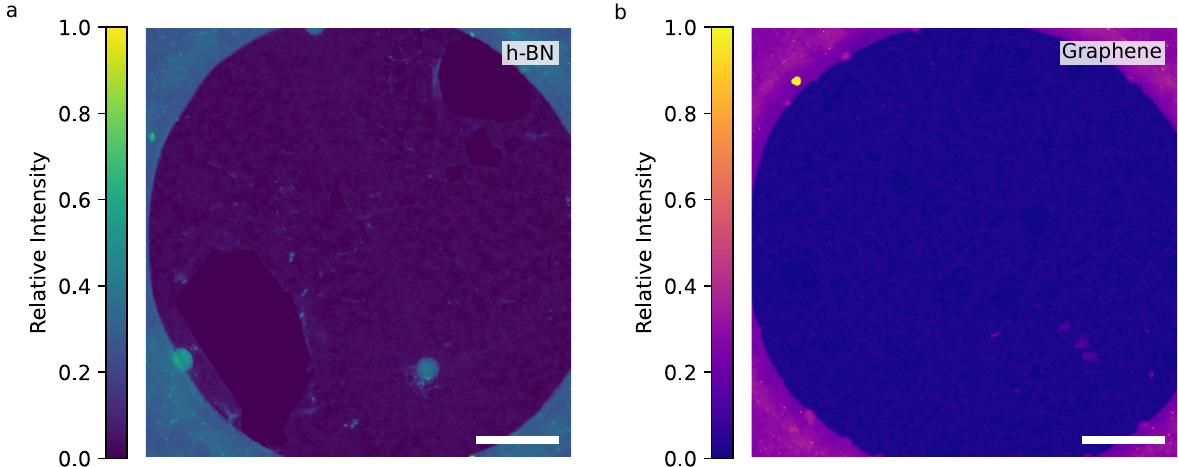}
\caption{{\bf Color maps used for \acrshort{hbn} and graphene.} Example false color MAADF-STEM images without contrast enhancement of suspended a) \acrshort{hbn} and b) graphene on the sample support (bright features in the corners), where the same color maps shown here were also applied to all other ADF-STEM images, both in the main text and the Supporting Information. There, the contrast was adjusted to highlight relevant features in the respective images. Scale bars: 200~nm.}
\label{fig:colorbar_hole_images}
\end{figure}

To improve the visual clarity and ease in comparability of the different sample types for the reader, false colors were applied to all ADF-STEM images of graphene and \acrshort{hbn} throughout the main text and the Supporting Information. Figure~\ref{fig:colorbar_hole_images} shows example images of graphene and \acrshort{hbn} suspended over the holey carbon sample support (bright intensities in the corners of the images) without any contrast enhancements applied. To highlight various features in other images presented in this work, different linear contrast enhancements were applied, while keeping the false color grading the same, and we refer the reader to the color bars presented in Figure~\ref{fig:colorbar_hole_images} for relative intensity comparisons.

\subsection*{Digital image processing}

\begin{figure}[!t]
\centering
\includegraphics[width=\textwidth]{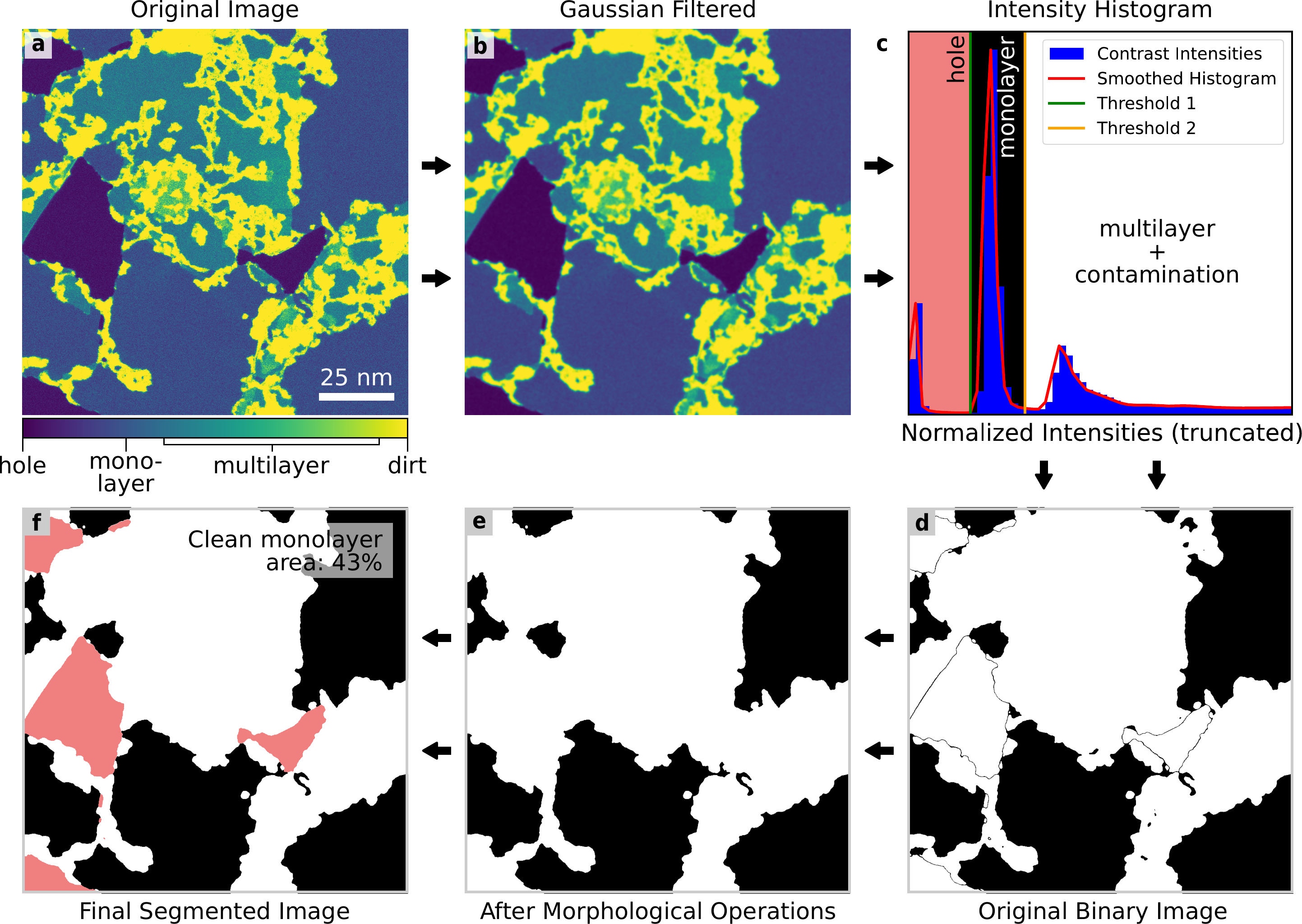}
\caption{{\bf Digital image processing steps for image thresholding and segmentation.} a) Example MAADF-STEM image used to identify clean monolayer coverage. This one contains holes (darkest contrast), clean monolayer \acrshort{hbn} (second darkest contrast) and bilayer structures, and contamination (brighter contrast). b) The same frame after applying a Gaussian filter (kernel size 17~px, $\sigma=2$~px). c) Contrast intensity histogram (1024 bins total) of the image in b). The histogram is truncated (56 bins are shown) to highlight relevant features. The red line resulted from iterative histogram smoothing. The colored vertical lines mark the two intensity thresholds. d) Original binary image obtained from the double thresholding process. Clean monolayer regions are black, other features are white. e) Binary image from panel d) after closing, opening, and inverted area opening (hole-filling) to remove artifacts from the thresholding process, and to remove small clean areas. f) Final segmented image, where clean monolayer areas are black, holes are pink, and multilayer regions and contamination are white.}
\label{fig:thresholding_demonstration}
\end{figure}

We identified monolayer graphene or \acrshort{hbn} as the darkest features in the image, or the second darkest features if the frame contains a hole. An example frame is shown in Figure~\ref{fig:thresholding_demonstration}a. To determine if a frame contains a hole, we employed an automated detection algorithm, where around ten reference images acquired during the same microscopy session that had been manually identified as containing pixels attributed to vacuum were selected as reference images. They were subjected to a Gaussian filter (kernel size 17×17~px, $\sigma=2$~px), from which a contrast intensity histogram (2048 bins) was generated. This histogram was iteratively smoothed using a one-dimensional Gaussian filter with a width of 3 bins ($\sigma=0.2$) for 50 iterations. Afterwards, the peak position of the leftmost peak (lower intensity values), which corresponds to the average intensity of the noise floor in a given frame, was identified and stored as reference data. The same procedure was then carried out for all frames acquired during the same microscopy session. If the leftmost peak of an image in its corresponding histogram was within the reference data's span or within $\pm$~3 standard deviations (whichever was greater), the image was labeled as containing a hole and subsequently added to the reference data before continuing with the next frame. This procedure was iteratively executed until no new frames were added to the reference images.

To determine the relative amount of clean monolayer area in each frame, a Gaussian filter (kernel size 17×17~px, $\sigma=2$~px) was first applied (Figure~\ref{fig:thresholding_demonstration}b) to aid in peak detection in the normalized contrast intensity histogram (1024 bins). The resulting histogram (Figure~\ref{fig:thresholding_demonstration}c, blue bars) was iteratively smoothed using a one-dimensional Gaussian filter with a width of 3 bins ($\sigma=0.2$) until the number of peaks reached a constant value. This smoothed histogram (Figure~\ref{fig:thresholding_demonstration}c, red line) was then thresholded using either the minimum or triangle method based on the number of identified peaks, with a difference between the two methods of typically less than 2\%. This way, the intensity peak corresponding to the clean monolayer regions in the image was reliably isolated (success rate $>95\%$) from the contamination and other higher-contrast features, such as multilayer regions, in addition to holes in the material, if applicable. The thresholds were used to generate a binary image, where the contaminated areas (and holes) were masked (Figure~\ref{fig:thresholding_demonstration}d). Subsequently, the binary image underwent a series of morphological operations, starting with an opening process with a circular kernel with a diameter of 0.5~nm (4~px), roughly corresponding to two lattice constants for both graphene and \acrshort{hbn}. This was done to both remove small clean areas and to separate barely connected ones (by contrast). If the image contained a hole, closing with the same kernel was performed before the opening process to close areas between the holes and the adjacent covered areas (thin black contours in Figure~\ref{fig:thresholding_demonstration}d), which had resulted from the Gaussian blur prior. Then, 8-connectivity inverted area opening (hole-filling) was performed to remove connected clean regions smaller than 16~nm$^2$ (1024~px$^2$), resulting in Figure~\ref{fig:thresholding_demonstration}e. This size threshold was chosen since areas as small as approximately 8~nm$^2$ have been utilized in recent literature to identify monolayer phosphorene\cite{speckmann_electron-beam-induced_2025} and single phosphorus dopants in graphene\cite{susi_single-atom_2017}. The chosen size was doubled from 8 to 16~nm$^2$ to account for differences in the morphology of the clean areas. Figure~\ref{fig:thresholding_demonstration}f shows the final segmented image, where the black pixels correspond to atomically clean areas, white pixels are contamination and multilayer structures, and pink regions are holes in the material. For each imaged hole in the sample support, the total number of black pixels and valid pixels (total area excluding holes and sample support) was summed up, with the clean area in $\%$ being calculated as the fraction of black pixels to all valid pixels.

\begin{figure}[!t]
\centering
\includegraphics[width=\textwidth]{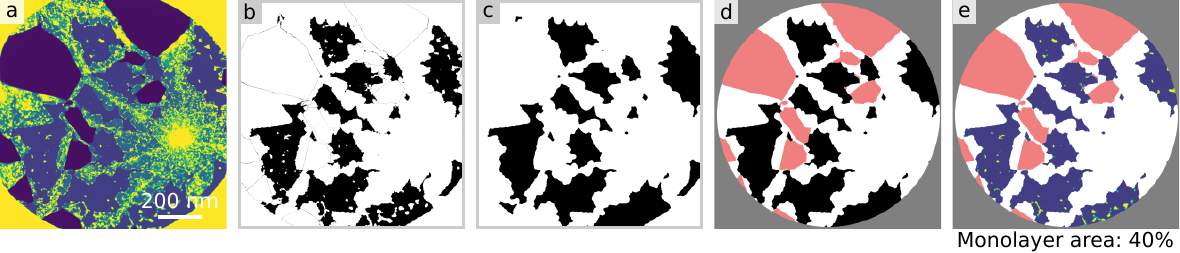}
\caption{{\bf Relative multilayer estimation of suspended h-BN.} a) MAADF-STEM image of a hole in the sample support. The darkest contrast is holes, the second darkest contrast is clean monolayer h-BN, and brighter features are multilayer structures and contamination. The bright features in the corners are the sample support. b) Original binary image after the thresholding process. Here, black pixels are clean monolayer regions, and white pixels are contamination, multilayer structures, and the sample support. c) Panel b) after five consecutive erosion and dilation steps, followed by area opening. Here, black pixels mark monolayer regions (regardless of contamination inside them). d) Final segmented image, where black areas are monolayer h-BN, pink areas are holes, white areas are multilayer structures, and the grey region masks the sample support. e) Mask obtained in panel d) overlaid on top of the original image in panel a), where only monolayer regions are not masked.}
\label{fig:multilayer_estimation_showcase}
\end{figure}

{\bf Monolayer area estimation}--- MAADF-STEM images of entire holes in the sample support (Figure~\ref{fig:multilayer_estimation_showcase}a) that visibly contain multilayer areas were analyzed to estimate the amount of exposed monolayer area for each imaged hole. They underwent an image thresholding and segmentation process analogous to the one outlined above. This way, the clean monolayer was separated from contamination and multilayer structures, as well as the sample support and holes in the material (Figure~\ref{fig:multilayer_estimation_showcase}b). This resulting binary image was then subjected to a series of four erosion processes, followed by four dilation processes, each with a circular 21×21~px kernel to remove contamination while retaining the general size and shape of (large) multilayer structures (Figure~\ref{fig:multilayer_estimation_showcase}c). Then, area opening was performed to remove connected contaminated areas that survived the previous process. The size threshold was varied between 2500~nm$^2$ and 6000~nm$^2$ depending on the frame. Finally, a circular mask with a diameter between 900~nm and 1000~nm (depending on the frame) was applied to mask areas containing the sample support (Figure~\ref{fig:multilayer_estimation_showcase}d). The monolayer area was then approximated by calculating the number of black pixels divided by the number of non-masked pixels. Figure~\ref{fig:multilayer_estimation_showcase}e shows the mask containing multilayer (white) and holes (pink) overlaid on top of the original image.

\subsection*{Mobile contamination}

\begin{figure}[!t]
\centering
\includegraphics{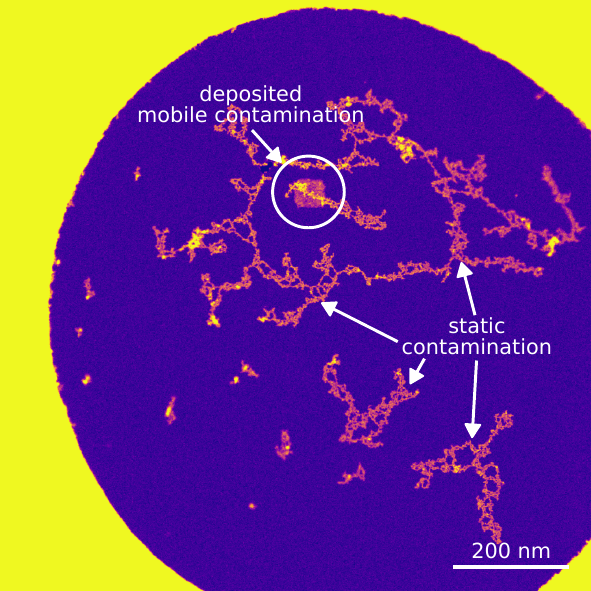}
\caption{{\bf E-beam-induced hydrocarbon deposition on graphene after longer UHV storage.} MAADF-STEM image of clean freestanding graphene (darkest contrast) with (static) dendritic contamination structures (brighter contrast) on its surface, where the visible square (white circle) originates from mobile contamination being pinned down due to e-beam-induced hydrocarbon deposition with the scanning probe.}
\label{fig:mobile_contamination}
\end{figure}

While we claim to have achieved large-area atomically clean graphene and h-BN, we cannot completely exclude the presence of hydrocarbon species diffusing on the surface ("mobile contamination"), as there have been reports of their existence even when the sample is kept at 800~°C during STEM imaging\cite{dyck_your_2025}. Their presence is difficult to directly image until they stick to the surface via e-beam-induced hydrocarbon deposition, where the hydrocarbon molecules are dissociated by the e-beam and then undergo a cross-linking process until they eventually stick to the surface. However, we did not observe this behavior even during extensive imaging on the same day the bake had been carried out, as can be seen in Supplementary Video 1 and Supplementary Video 2 (\SI{4}{\micro\second} pixel time, 130~pA beam current, 32x32~nm$^2$ nominal FOV, 512x512~px). However, after storage in near-UHV conditions for over three weeks, diffuse hydrocarbon molecules were almost immediately deposited onto the surface, even after only brief e-beam exposure (Figure~\ref{fig:mobile_contamination}), making it practically impossible to image what had previously been identified as atomically clean graphene. It is possible that, immediately after the heating, the residual heat on the sample and sample holder reduced the adsorption of mobile contamination onto them due to their higher temperature compared to the sample stage, a difference that is equalized after longer storage. It may also be the case that contaminant species inside the near-UHV sample storage atmosphere eventually deposit onto the surface, in addition to the diffusion of surface-bound hydrocarbon molecules on the surfaces inside the sample storage chamber. Nevertheless, given that no e-beam-induced hydrocarbon deposition was observed during the initial data collection, we have high confidence that the potential presence of mobile contamination did not affect the measured clean area.

\subsection*{Sample support damage}

\begin{figure}[!t]
\centering
\includegraphics[width=\textwidth]{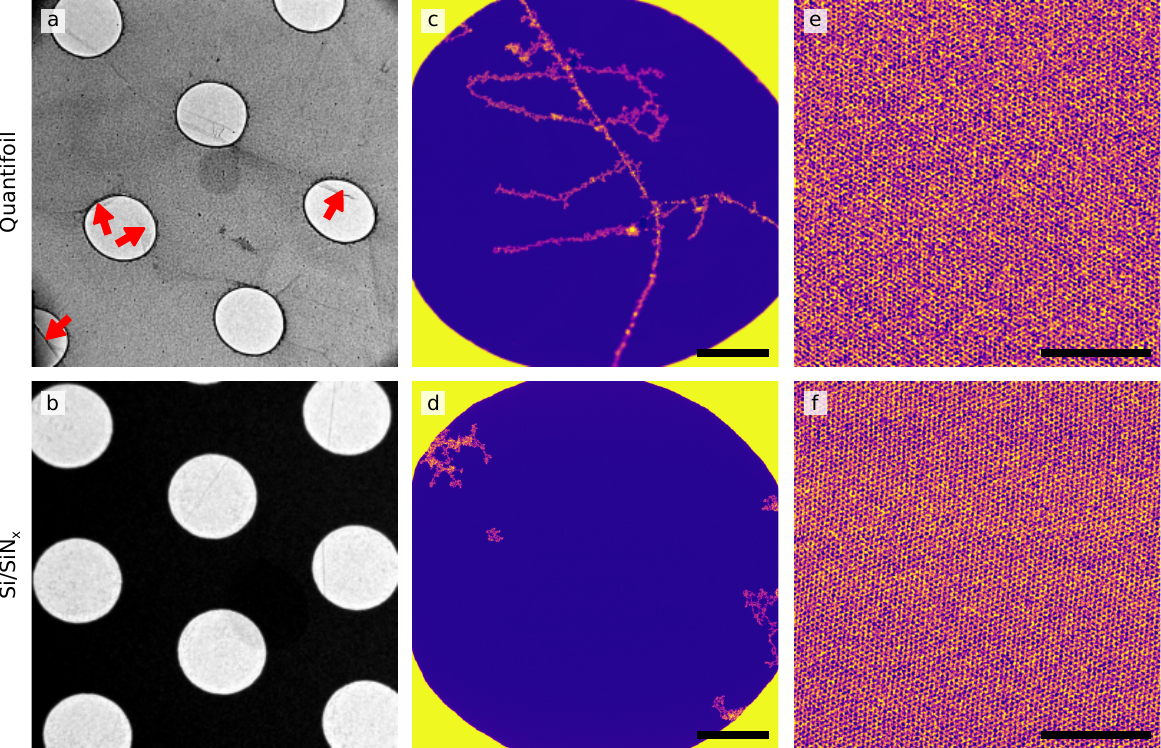}
\caption{{\bf Differences in thermal response of the sample support.} Bright field Ronchigram image of graphene a) on a Quantifoil (QF) holey carbon support film and b) on a perforated Si/SiN$_x$ support after annealing at 450~°C for 3~h. Nominal hole diameters \SI{1}{\micro\meter}. The red arrows in a) point towards damage in the graphene due to sample support warping. MAADF-STEM images showing a typical hole containing clean suspended graphene c) on QF and d) on Si/SiN$_x$, with higher magnification MAADF-STEM images in e) and f) showing large areas of pristine clean graphene on QF, and Si/SiN$_x$, respectively. Scale bars: c, d) 200~nm, e, f) 5~nm.}
\label{fig:qf_warping}
\end{figure}

At higher annealing temperatures, we observe substantial warping and damage of the structural sample support, likely occurring from the mismatch of the thermal expansion coefficient of the gold bars and amorphous carbon film, in addition to thinning of the already thin (ca. 10~nm) amorphous carbon support membrane. Figure~\ref{fig:qf_warping}a shows the observed compressions of the imaged projections of the holes in the support film, which manifest as elliptic rather than circular appearing holes, originating from thermally-induced stretching of the sample support. This not only causes mechanical strain, creating tears and cracks in the graphene (red arrows in Figure~\ref{fig:qf_warping}a), but also makes it more difficult to get larger areas into focus at once with the scanning probe due to strong height differences within each hole (Figure~\ref{fig:qf_warping}b), in addition to support-induced vibrations which manifest in non-ideal imaging conditions, limiting acquisition of atomically-resolved images. Therefore, another graphene sample prepared via electrochemical delamination was transferred onto a holey silicon nitride support membrane with a rigid 200~nm membrane thickness and was heated at 450~°C for 2~h in UHV. After annealing at 450~°C, the sample support was still in pristine condition (Figure~\ref{fig:qf_warping}c), mitigating support-induced corrugations of graphene (Figure~\ref{fig:qf_warping}d) and thus also minimizing strain-induced damage to the material. This sample exhibited a similar level of cleanliness as the one on the holey carbon support film that had also been annealed at 450~°C, with a measured average clean area above 95\% (Figure~\ref{fig:temperature_comparison_graphene_new}a, main text, purple dots), slightly higher than the 90\% measured from the other sample support. However, we attribute this minimal difference to local differences from the source material and transfer, in addition to contamination overestimation arising from unfocused features caused by the earlier-mentioned height variations in the other samples, rather than due to a different thermal response of the Si/SiN$_x$ compared to the holey amorphous carbon support.

\newpage

\subsection*{Other trace contaminants}

\begin{figure}[!t]
\centering
\includegraphics[width=\textwidth]{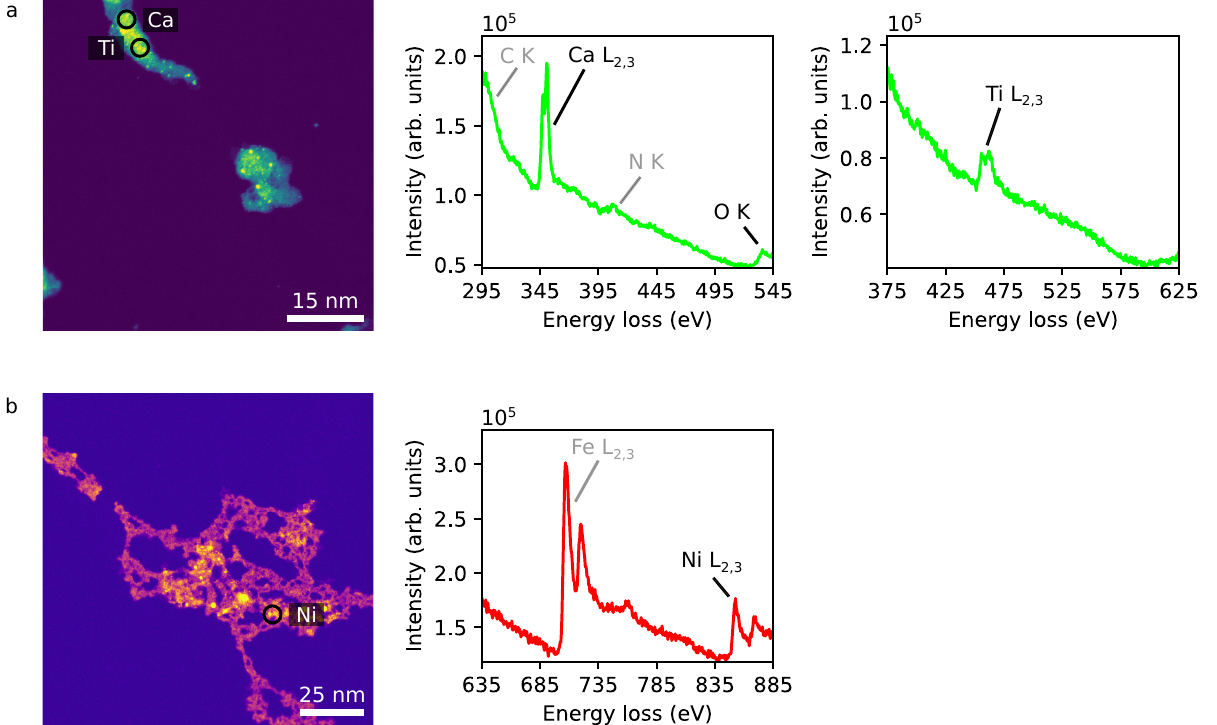}
\caption{{\bf Identified trace contaminants.} a) Ca L$_{2,3}$ (and O K) and Ti L$_{2,3}$ EEL spectra (not background-subtracted) acquired on contaminated sites (marked in the accompanying MAADF-STEM image) on \acrshort{hbn} after annealing at 550~°C. b) Ni L$_{2,3}$ spectrum (not background-subtracted) acquired within metallic contamination (marked in the accompanying MAADF-STEM image) on graphene after annealing at 450~°C. Core loss edges of other contaminant species (C, Fe) and N from the material are also visible.}
\label{fig:trace_contaminants_eels}
\end{figure}

Among the ubiquitous hydrocarbon contamination, as well as the Cu and Fe nanoclusters discussed in the main text, we could also identify other contaminants (Figure~\ref{fig:trace_contaminants_eels}), namely oxygen, calcium, titanium, and nickel, though rarely. Detecting oxygen was challenging, as we found it not only not to be abundant, particularly after the annealing step, but also due to its decreasing signal strength with prolonged electron beam exposure\cite{leuthner_scanning_2019} during spectrum acquisition, which would be favorable to obtain a better signal-to-noise ratio. However, we could identify it alongside calcium on one of our samples (\acrshort{hbn} annealed at 550~°C), with Ca likely originating from impure deionized water used during the sample transfer process. The concurrent presence of Ca, O, and C suggests calcium oxide or calcium carbonate as the contaminant. We could also identify titanium on some of our samples, in particular ones that have been stored in \acrshort{uhv} for longer, likely originating from Ti sputtering from the ion pumps in the vacuum system. Nickel could also be found on one of our graphene samples (annealed at 450~°C) alongside a Fe nanocluster, which could originate from any of the stainless steel parts of the vacuum system, or could have evaporated from the K-type (Ni-Cr/Ni-Al) thermocouple inside the \acrshort{uhv} heating chamber during the annealing procedure. If Ni had indeed evaporated from the thermocouple, one would also expect Cr deposition on the sample surface. However, it was not detected on any of the samples analyzed by EELS, although its presence cannot be ruled out without conducting a more thorough investigation.

\subsection*{UHV heating chamber design}

\begin{figure}[hb]
\centering
\includegraphics[width=\textwidth]{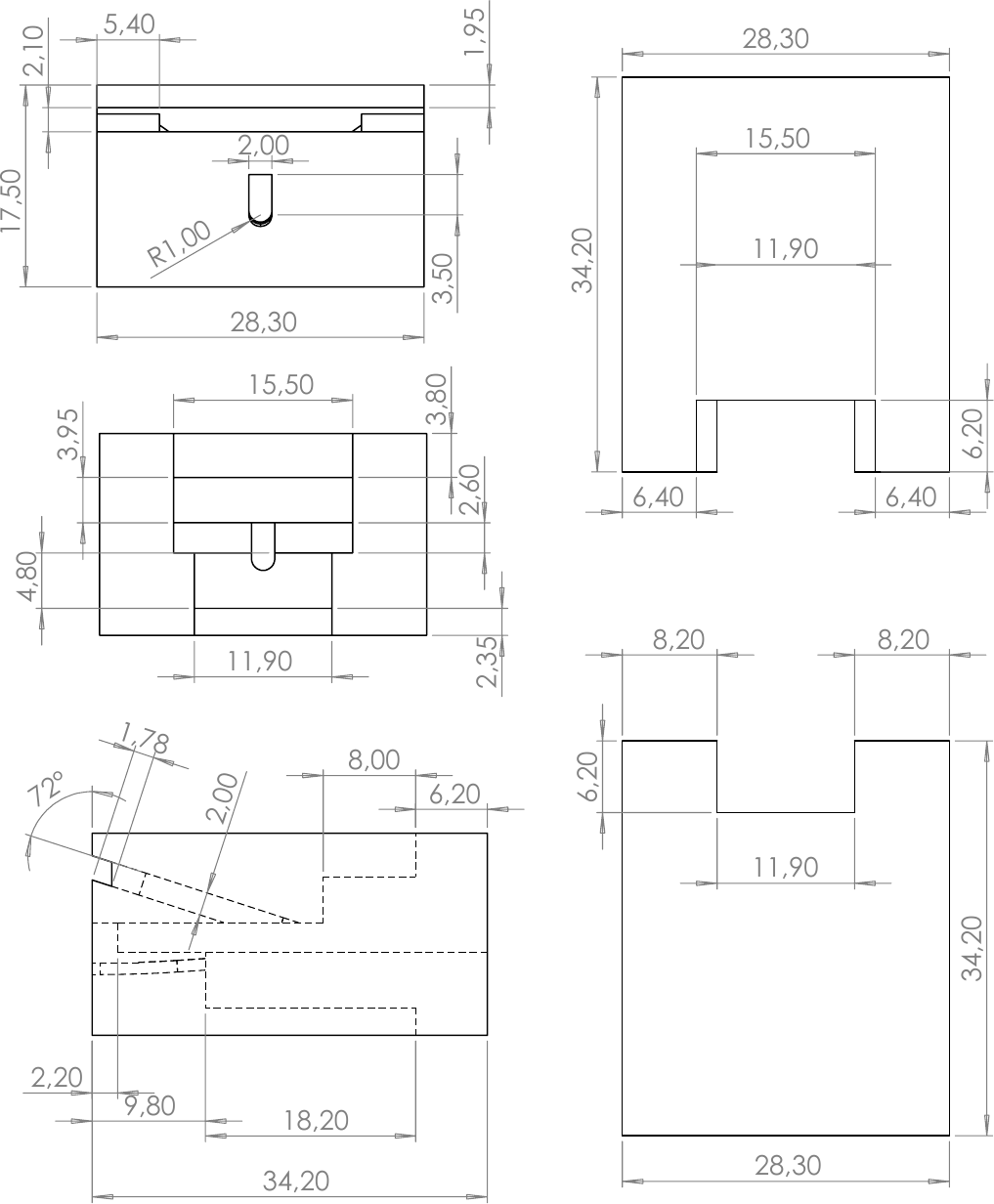}
\caption{{\bf Construction plan for the 3D-printed alumina furnace.} Back, front, side, top, and bottom view orientations of the alumina furnace, with measurements of distances, angles, and hole radii are displayed. All lengths in mm.}
\label{fig:sw_aloxide_oven}
\end{figure}

\begin{figure}[hb]
\centering
\includegraphics{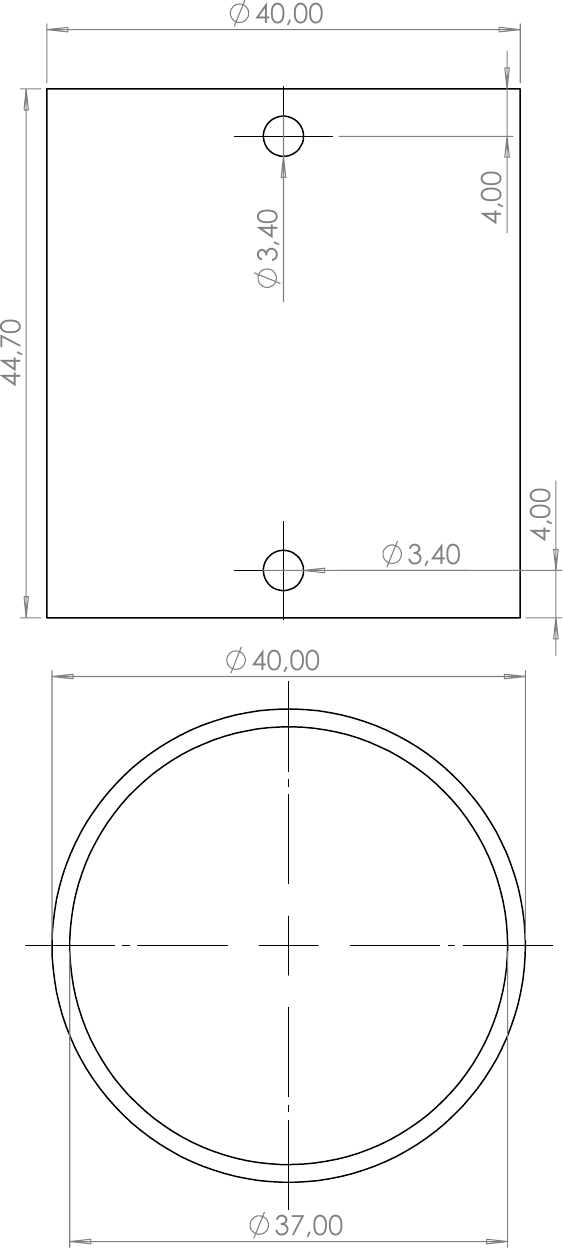}
\caption{{\bf Construction plan for the cylindrical stainless steel frame of the oven.} Top and front view orientations of the alumina furnace, with measurements of distances and hole radii are displayed. All lengths in mm.}
\label{fig:sw_rohr}
\end{figure}

\begin{figure}[hb]
\centering
\includegraphics{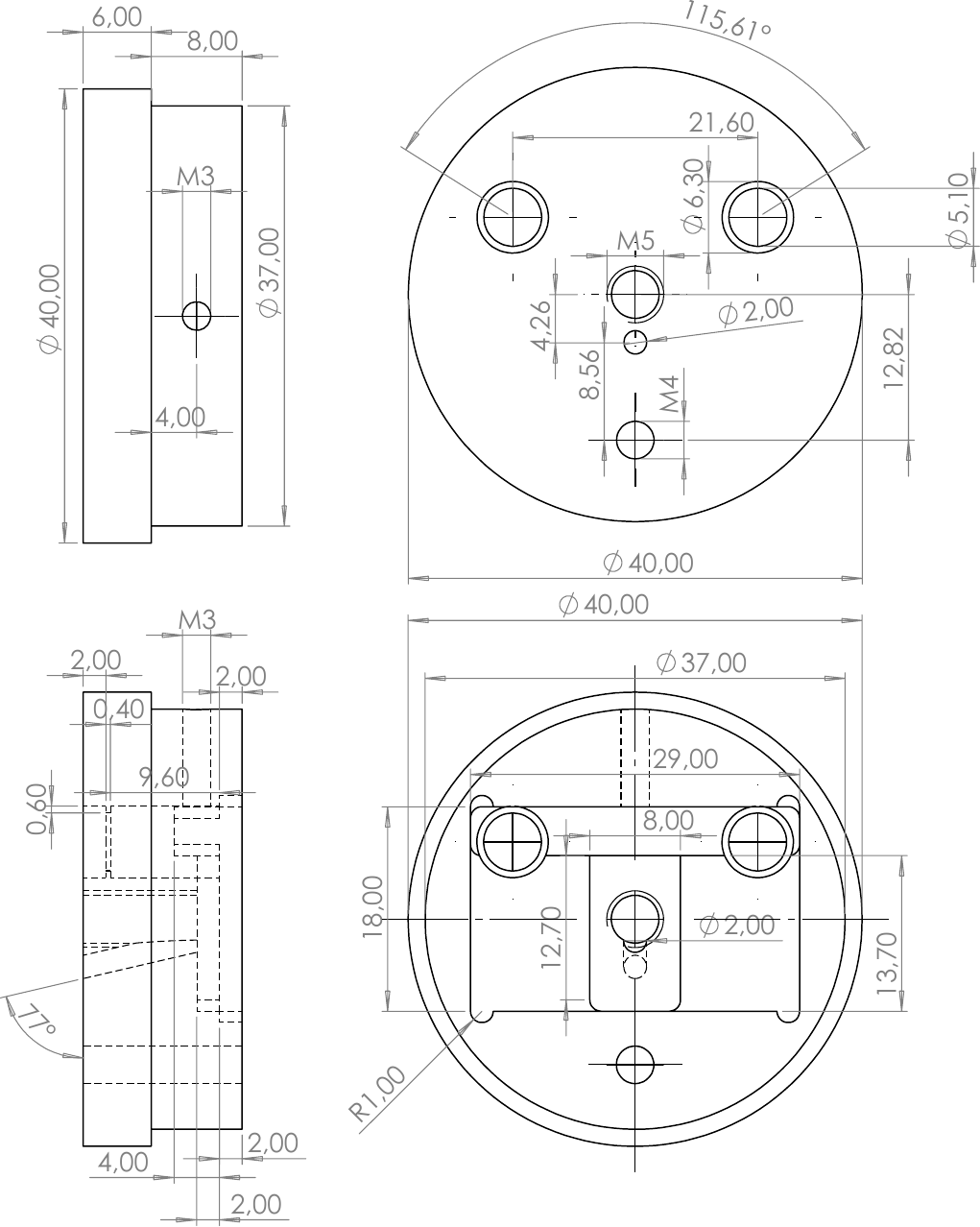}
\caption{{\bf Construction plan for the backside of the oven.} Top, side, front, and back view orientations of the backside of the oven frame, with measurements of distances, angles, and hole diameter, are displayed. ISO metric screw threads (M3, M4, M5) are used. All lengths in mm.}
\label{fig:sw_boden_neu}
\end{figure}

\begin{figure}[hb]
\centering
\includegraphics{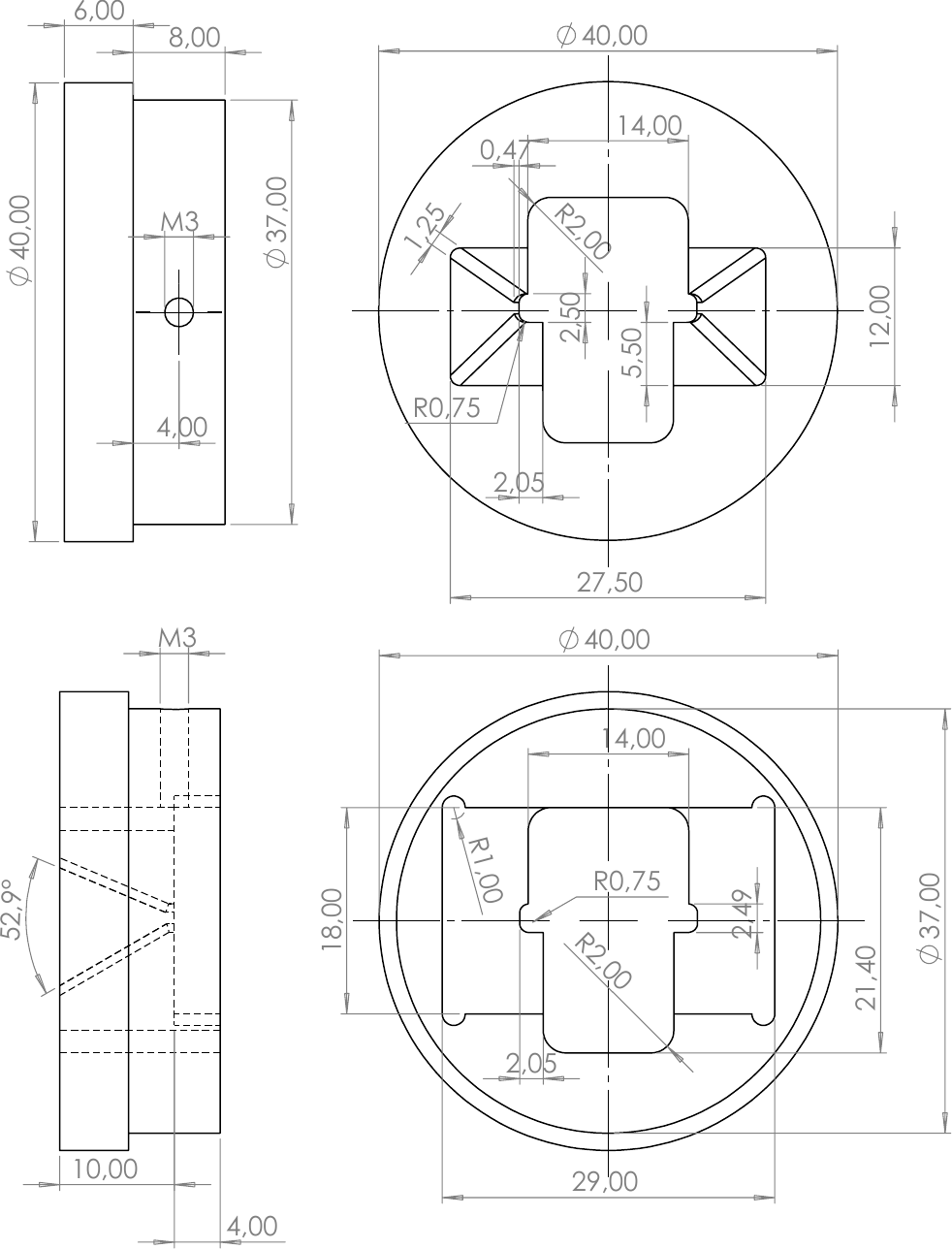}
\caption{{\bf Construction plan for the frontside of the oven.} Top, side, front, and back view orientations of the backside of the oven frame, with measurements of distances, radii, fillet radii, and angles, are displayed. ISO metric screw threads (M3) are used. All lengths in mm.}
\label{fig:sw_deckel}
\end{figure}

\begin{figure}[hb]
\centering
\includegraphics{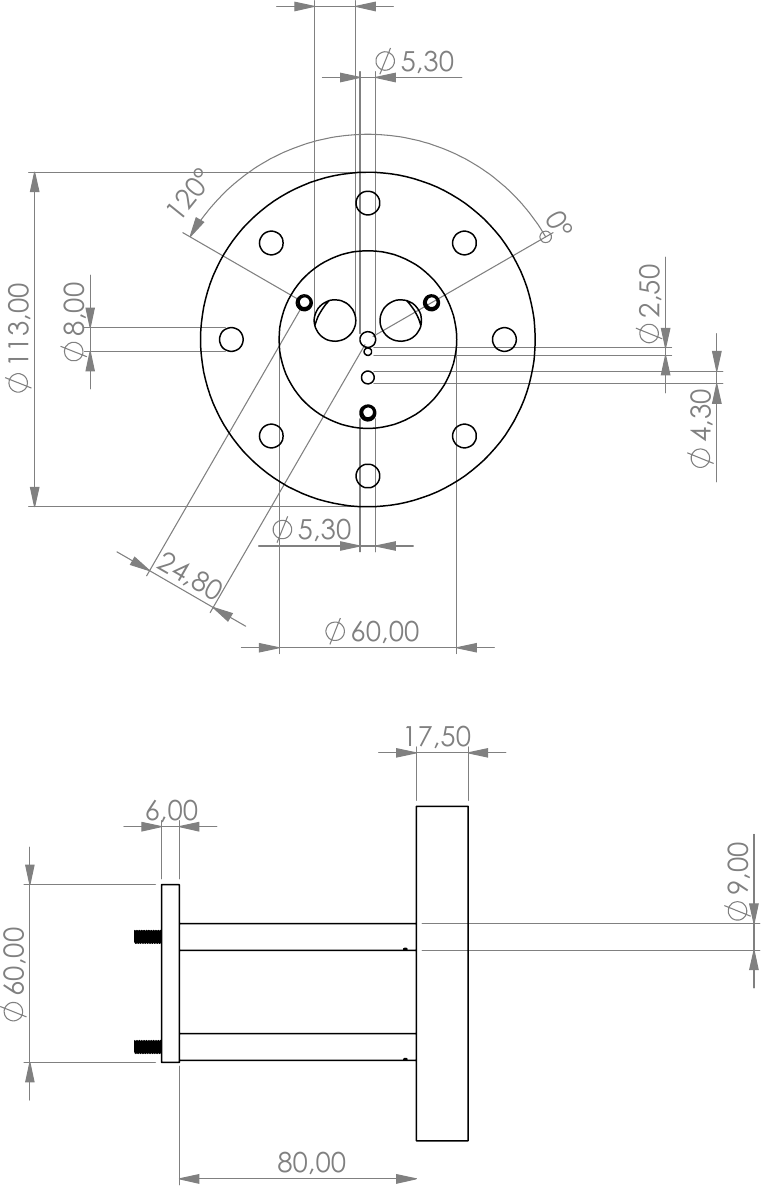}
\caption{{\bf Construction plan for the UHV heating chamber fixture assembly.} Front and side view orientations of the backside of the oven frame, with measurements of distances, angles, and hole radii, are displayed. All lengths in mm.}
\label{fig:sw_halterung}
\end{figure}

The UHV heating chamber consists of a custom 3D-printed alumina furnace (Figure~\ref{fig:sw_aloxide_oven}) that accommodates the titanium sample holder, as well as a \SI{0.5}{\milli\meter} thin tungsten filament resting in a 2~mm thick cavity responsible for the resistive heating process. The filament is placed just out of line of sight of the sample to avoid locally high temperatures and allow a more accurate temperature measurement using an ungrounded Inconel 600-sheathed and mineral-insulated K-type thermocouple, which is positioned approximately \SI{1}{\milli\meter} away from the sample and is in physical contact with the sample holder. The oven is encompassed in a cylindrical stainless steel frame (Figure~\ref{fig:sw_rohr}) with holes on one of the flat sides of the cylinder (Figure~\ref{fig:sw_boden_neu}) to attach the control electronics and temperature probe, and a sample holder entry on the opposing flat side of the cylinder (Figure~\ref{fig:sw_deckel}). The whole assembly is mounted to a zero-length reducing flange at the back of the chamber through stainless steel rods (Figure~\ref{fig:sw_halterung}), and is integrated within the interconnected \acrshort{uhv} system CANVAS at the University of Vienna\cite{mangler_materials_2022}.

\begin{figure}[hb]
\centering
\includegraphics[width=\textwidth]{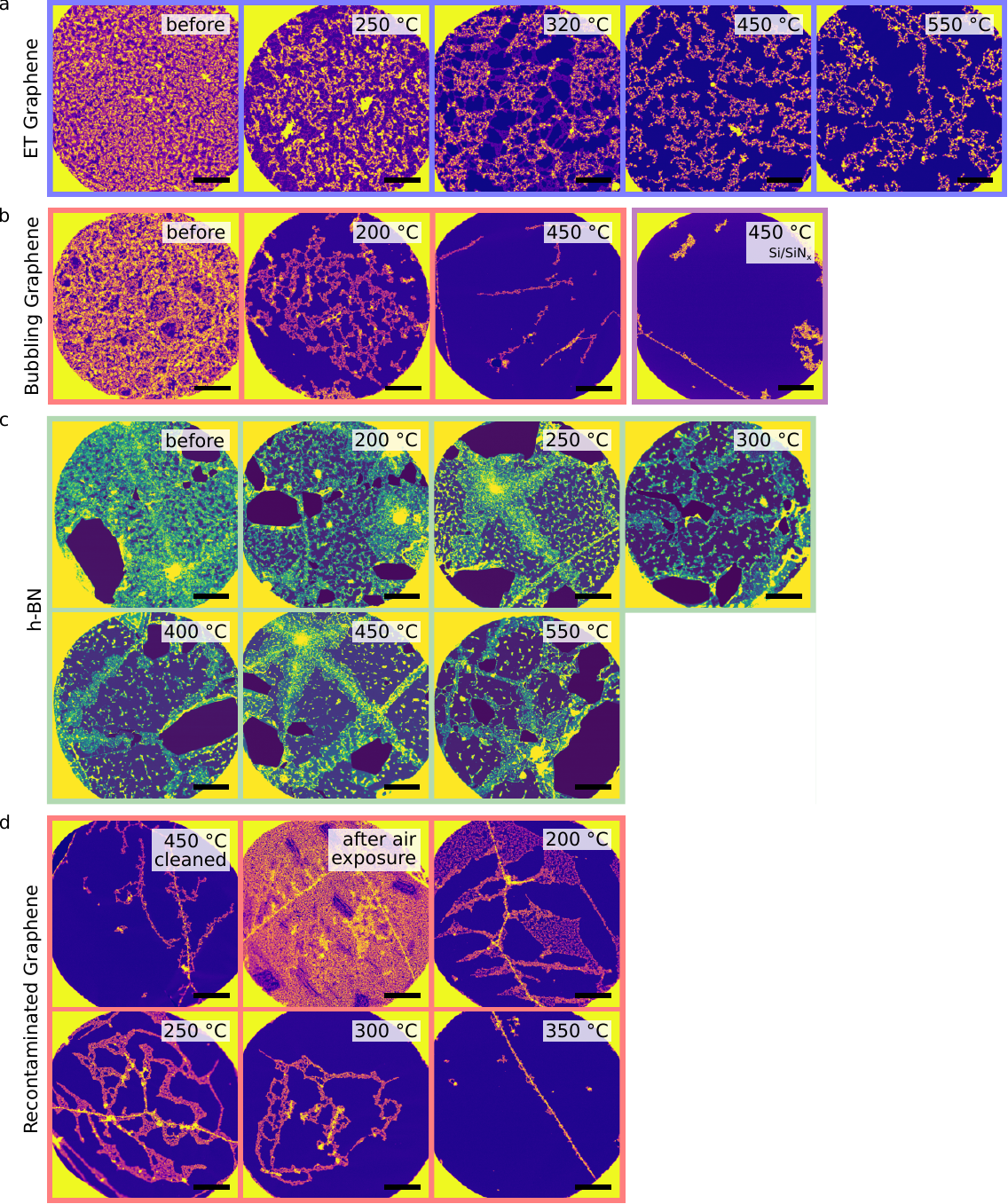}
\caption{{\bf MAADF-STEM images of non-cropped frames used for temperature comparison.} Original (non-cropped) images of the cropped MAADF-STEM images shown in a,b) Figure~\ref{fig:temperature_comparison_graphene_new}, c) Figure~\ref{fig:temperature_comparison_hbn}, and d) Figure~\ref{fig:recontaminated_cleaning} (main text). Scale bars: 200~nm.}
\label{fig:non-cropped_temperature_comparison}
\end{figure}

\begin{figure}[hb]
\centering
\includegraphics[width=\textwidth]{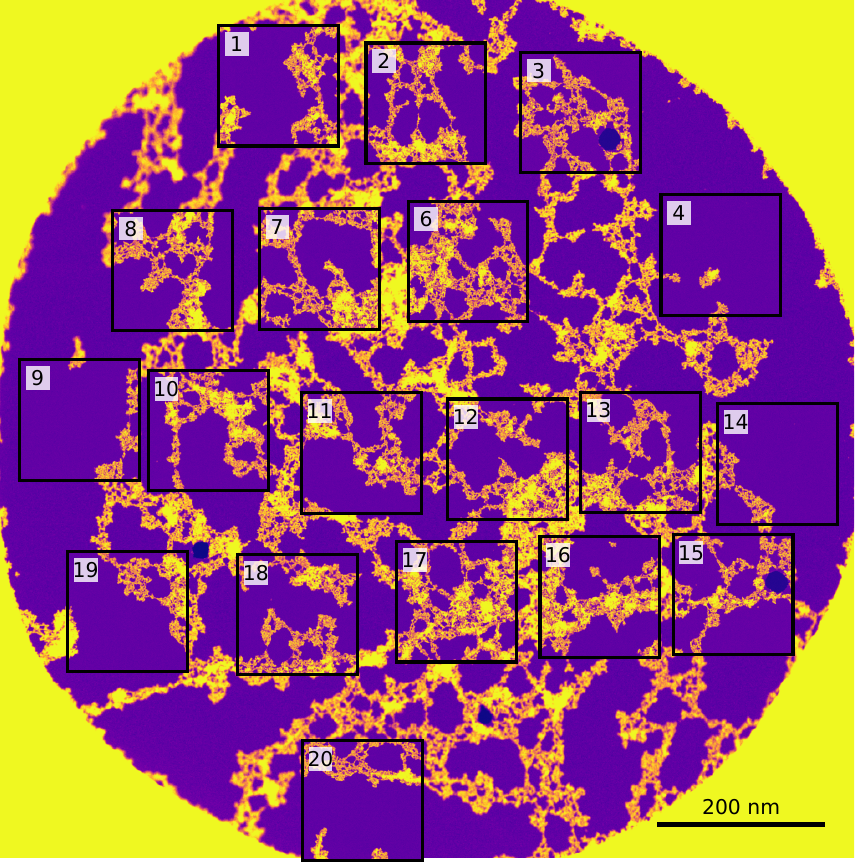}
\caption{{\bf Depiction of the image acquisition process within one example hole in the sample support.} MAADF-STEM images of monolayer graphene showing the positions of acquired frames within a hole in the sample support. The darkest contrast is holes in the material, corresponding to vacuum intensity (see, for example, frames 3 and 15), the second darkest contrast is clean monolayer graphene. Individually acquired frames (nominal FOV 128x128~nm$^2$) are overlaid on top of an overview image of the entire hole in the background. The numbers in the top left corners of the insets indicate the respective frames in the hole, listed in the order of acquisition. Here, frame 5 was removed due to significant overlap with frame 4.}
\label{fig:all_frames_from_one_hole2}
\end{figure}

\begin{figure}[hb]
\centering
\includegraphics[width=\textwidth]{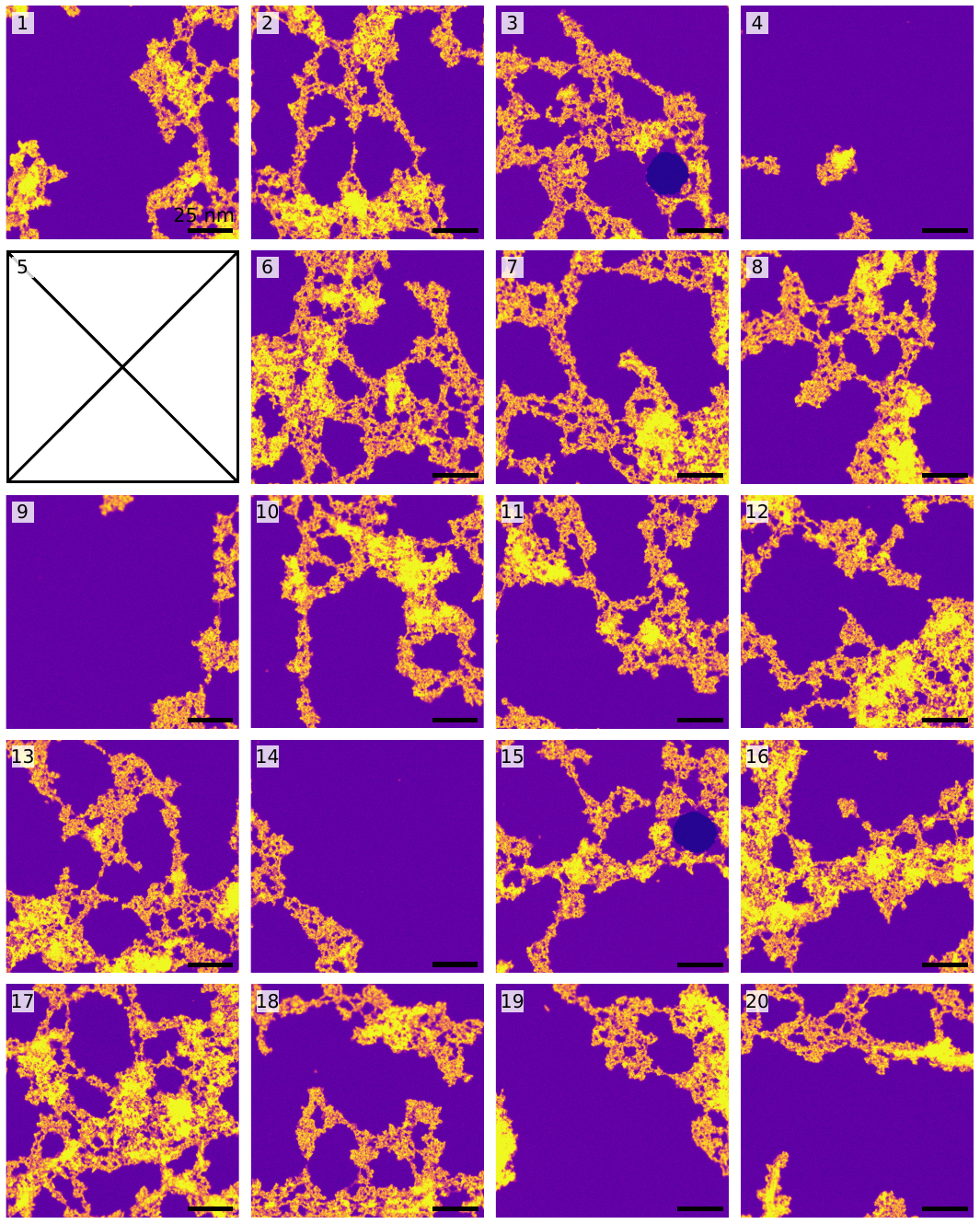}
\caption{{\bf MAADF-STEM images of all frames of one example hole for graphene (see Figure~\ref{fig:all_frames_from_one_hole2}.} The numbers in the top left corners of the other images shown here correspond to the respective frames acquired in the hole. In all images, the darkest contrast is that of clean monolayer graphene, except for frames 3 and 15, where the darkest contrast corresponds to holes in the material and represents the vacuum intensity. For the calculation of the relative clean area, the number of pixels of all frames acquired within one hole is summed up (see Digital image processing, Supporting Information). Here, 65\% of the total covered area is clean monolayer graphene.  Here, frame 5 was removed due to significant overlap with frame 4.}
\label{fig:all_frames_from_one_hole1}
\end{figure}

\begin{figure}[hb]
\centering
\includegraphics[width=\textwidth]{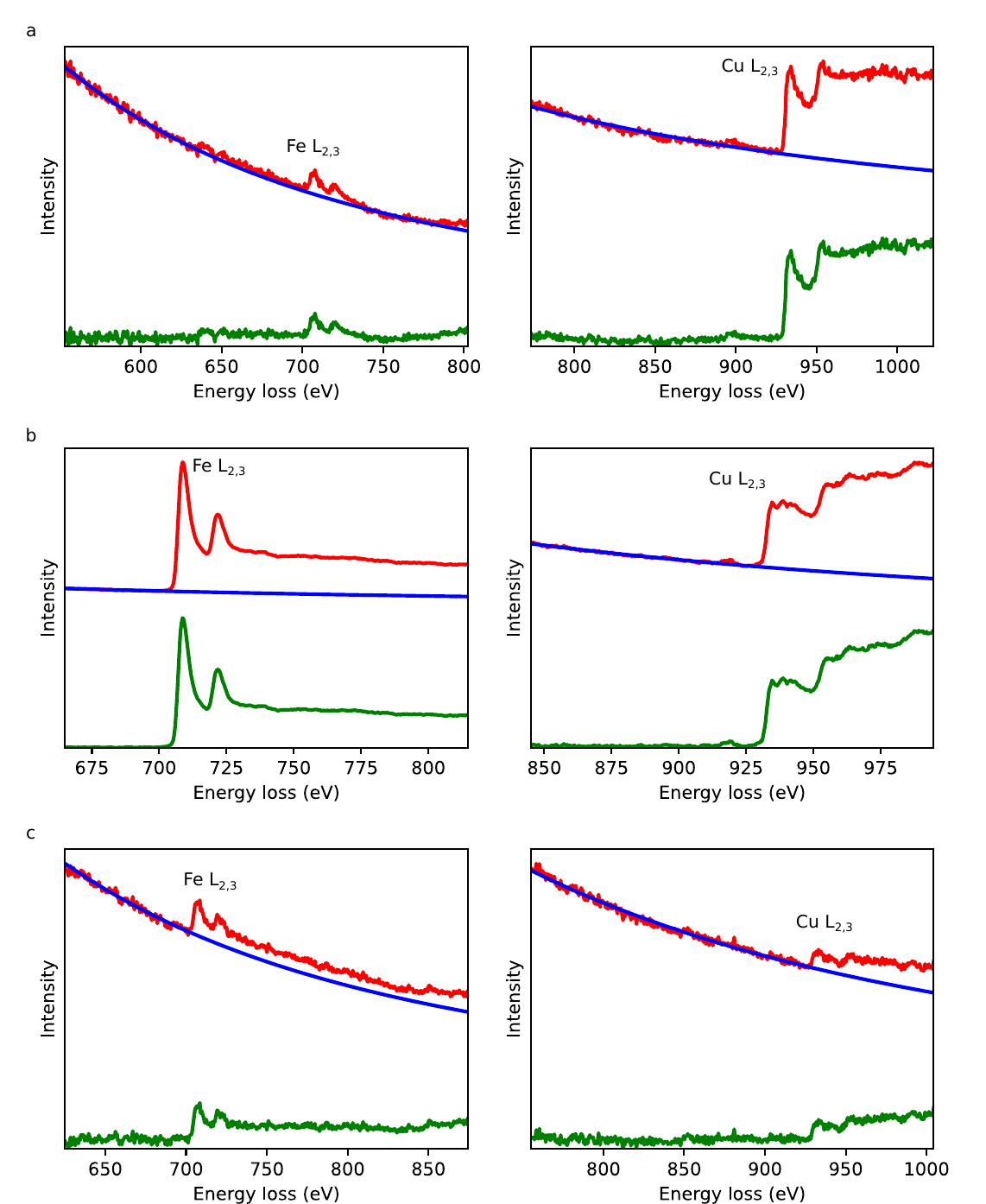}
\caption{{\bf Original EEL spectra including background subtraction.} Original Fe L$_{2,3}$ and Cu L$_{2,3}$ spectra (red) with added vertical offset, power law background subtraction function (blue), and resulting background-subtracted spectra (green) acquired on iron and copper clusters a) on h-BN (Figure~\ref{fig:metal_EELS_spectra_after_new}a), b) on Easy Transfer Graphene (Figure~\ref{fig:metal_EELS_spectra_after_new}b), c) on graphene transferred via electrochemical delamination (Figure~\ref{fig:metal_EELS_spectra_after_new}c) after annealing at 450~°C for 3~h.}
\label{fig:EELS_raw_data_metal_clusters}
\end{figure}

\begin{figure}[hb]
\centering
\includegraphics[width=\textwidth]{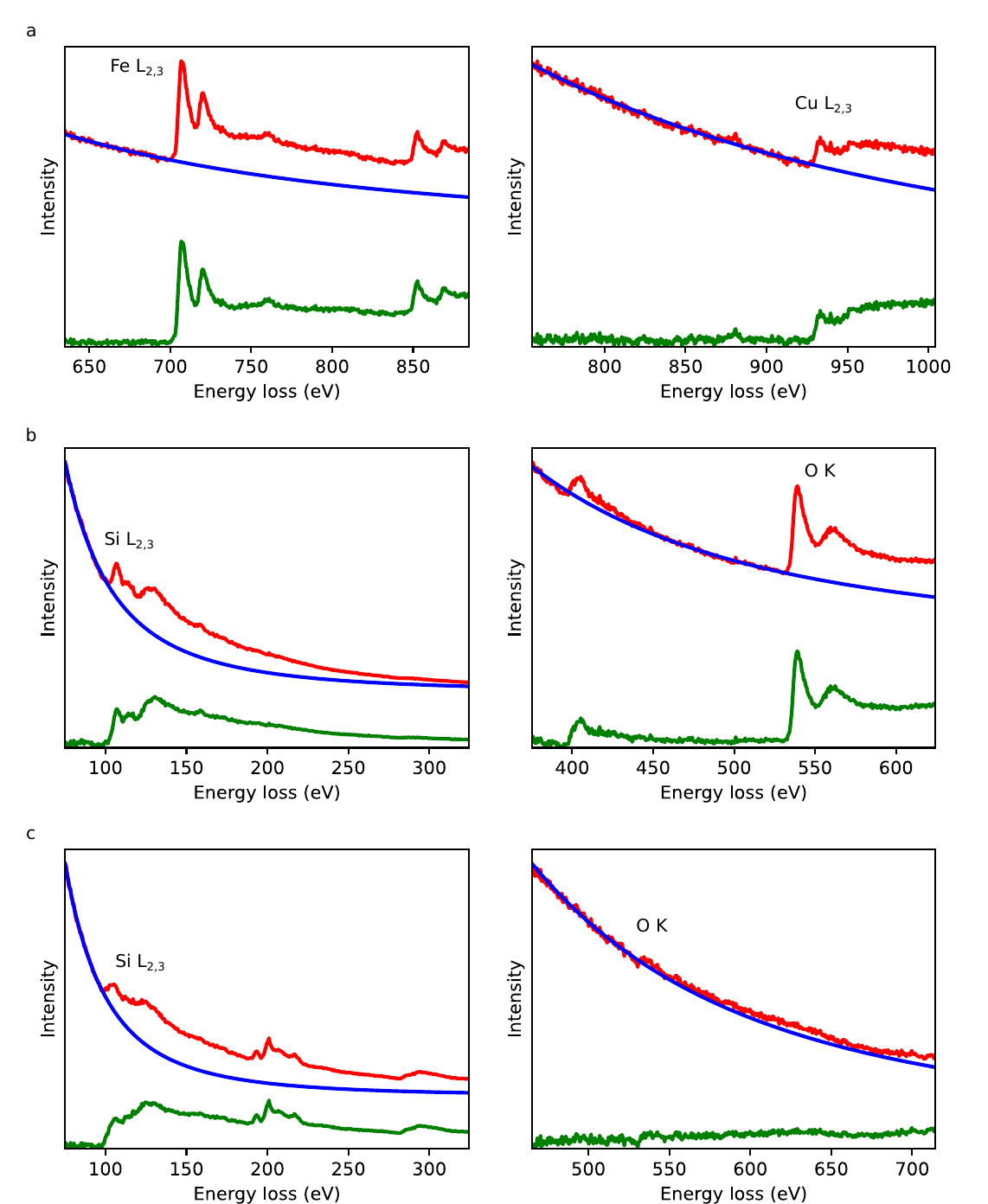}
\caption{{\bf Original EEL spectra including background subtraction.} Original spectra (red) with added vertical offset, power law background subtraction function (blue), and resulting background-subtracted spectra (green) acquired a) on iron and copper clusters within agglomerations of metal clusters on graphene transferred via electrochemical delamination after annealing at 450°C for 3~h (Figure~\ref{fig:graphene_cleaning_limitations_new}), b) on a SiO$_2$ nanocluster on h-BN before annealing, c) on hydrocarbon contamination next to a SiO$_2$ nanocluster on h-BN before annealing (Figure~\ref{fig:SiO2_EELS_multilayer_new}c). Clearly visible are also the a) Ni L$_{2,3}$, b) N K, and c) B K, and C K core loss edges.}
\label{fig:EELS_raw_data_cleaning_limitations}
\end{figure}

\printbibliography
\end{refsection}


\end{document}